\documentclass[10pt,journal]{IEEEtran}
\ifCLASSOPTIONcompsoc
\else
\fi
\ifCLASSINFOpdf
\else
 
\fi

\usepackage{booktabs} 
\usepackage[noadjust]{cite}
\usepackage{amsmath}
\usepackage{amssymb}
\usepackage{amsfonts}
\usepackage{float}
\usepackage{graphicx}
\usepackage{xcolor}
\usepackage{colortbl}
\usepackage{multirow}
\usepackage{mathtools}
\usepackage{algorithm}
\usepackage{algorithmicx}
\usepackage{algpseudocode}
\usepackage{tabto}
\usepackage{hhline}
\usepackage{enumitem}
\usepackage[font=small,labelfont={it}]{caption}
\usepackage[none]{hyphenat}
\usepackage{changepage}
\usepackage{pifont}

\algnewcommand\algorithmicinput{\textbf{Input:}}
\algnewcommand\Input{\item[\algorithmicinput]}

\algnewcommand\algorithmicoutput{\textbf{Output:}}
\algnewcommand\Output{\item[\algorithmicoutput]}

\makeatletter
\renewcommand\footnoterule{%
  \kern-3\p@
  \hrule\@width 0.5\columnwidth
  \kern2.6\p@}
  \makeatother

\begin{document}
\title{\fontsize{23}{30}\selectfont ForASec: \underline{For}mal \underline{A}nalysis of Hardware Trojan-based \underline{Sec}urity Vulnerabilities in Sequential Circuits}
\author{
    Faiq~Khalid,~\IEEEmembership{Member,~IEEE,} 
    Imran Hafeez Abbassi, Semeen Rehman, 
    Awais Mehmood Kamboh, \IEEEmembership{Senior~Member,~IEEE}, 
    Osman Hasan,~\IEEEmembership{Senior~Member,~IEEE}
    and~Muhammad Shafique,~\IEEEmembership{Senior~Member,~IEEE}\vspace{-25pt}
\IEEEcompsocitemizethanks{
    \IEEEcompsocthanksitem F. Khalid and S. Rehman, are with Technische Universit{\"a}t Wien (TU Wien), Vienna, Austria. E-mail: \{faiq.khalid, semeen.rehman\}@tuwien.ac.at.\protect
    \IEEEcompsocthanksitem I. H. Abbassi is with College of Aeronautical Engineering, NUST, Pakistan. E-mail: imran.abbasi@cae.nust.edu.pk \protect
    \IEEEcompsocthanksitem O. Hasan is with School of Electrical Engineering and Computer Science, National University of Sciences \& Technology, Islamabad, Pakistan. E-mail: osman.hasan@seecs.edu.pk.\protect
    \IEEEcompsocthanksitem A. M. Kamboh is with Department of Computer and Network Engineering, College of Computer Science and Engineering (CCSE), University of Jeddah, Jeddah 21589, Saudi Arabia and School of Electrical Engineering and Computer Sciences, NUST, Islamabad, Pakistan. E-mail: awais.kamboh@seecs.edu.pk; amkamboh@uj.edu.sa \protect
    \IEEEcompsocthanksitem M. Shafique is with Division of Engineering, New York University Abu Dhabi (NYU AD), Abu Dhabi, United Arab Emirates. E-mail: muhammad.shafique@nyu.edu \protect\\
}
\thanks{This work is supported in parts by the Austrian Research Promotion Agency (FFG) and the Austrian Federal Ministry for Transport, Innovation, and Technology (BMVIT) under the “ICT of the Future” project, IoT4CPS: Trustworthy IoT for Cyber-Physical Systems.}
}
\markboth{IEEE Transactions on Computer-Aided Design of Integrated Circuits and Systems}%
{F. Khalid \MakeLowercase{\textit{et al.}}: ForASec}
\maketitle
\begin{abstract}
In this paper, we propose a novel model checking-based methodology that analyzes the Hardware Trojan (HT)-based security vulnerabilities in sequential circuits with 100\% coverage while addressing the state-space explosion issue and completeness issue. In this work, the state-space explosion issue is addressed by efficiently partitioning the larger state-space into corresponding smaller state-spaces to enable distributed HT-based security analysis of complex sequential circuits. We analyze multiple ISCAS89 and trust-hub benchmarks for different ASIC technologies, i.e., 65nm, 45nm, and 22nm, to demonstrate the efficacy of our framework in identifying HT-based security vulnerabilities. The experimental results show that ForASec successfully performs the complete analysis of the given complex and large sequential circuits and provides approximately 6x to 10x speedup in analysis time compared to state-of-the-art model checking-based techniques.
\end{abstract}

\begin{IEEEkeywords}
Formal Verification, Hardware Trojans, Model Checking, Security Vulnerabilities, Gate-level modeling.\vspace{-5pt}
\end{IEEEkeywords}



\section{Introduction}
The increasing trend of outsourcing fabrication and split manufacturing to un-trusted foundries and manufacturing plants have imposed significant threats like hardware Trojans (HT) to the security of critical applications~\cite{li2016survey,bhasin2015survey,huang2020survey}. Therefore, it is imperative to develop HT detection (like propagation delay~\cite{farahmandi2020hardware,tran2019hardware,hoang2019hardware,vakil2020lasca}, temperature~\cite{yang2020hardware,tang2019activity} and power consumption~\cite{lodhi2017power,hoque2017golden,lodhi2016self,xue2017power,xue2017self} based techniques) and prevention techniques for designing robust systems that can tolerate HTs. Towards this goal, several HT detection and prevention techniques have been proposed. Though these techniques can detect or make circuits capable of handling the HTs, designing these techniques depends on the security vulnerability analyses. 

\begin{figure*}[!t]
	\centering
	\includegraphics[width=1\linewidth]{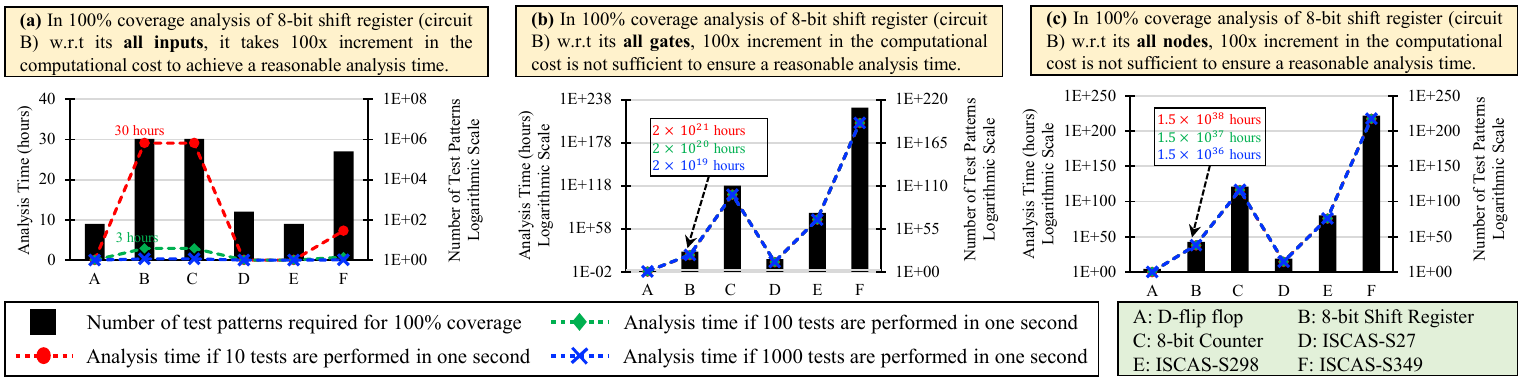} 
	\caption{\textit{Required number of test patterns and time for 100\% coverage for HT-based security vulnerability analysis of different sequential circuits (DFF, 8-bit shift register, 8-bit counter, and some of the ISCAS89 benchmarks, i.e., s27, s298, s349, s35932, s38417, and s35854) with respect to all inputs, all gates and all nodes. The analysis times are estimated as: for all inputs, $T_{inputs}= TP_{inputs}/n_{test}$, for all nodes, $T_{nodes} = TP_{nodes}/n_{test}$, and for all gates $T_{gates} = TP_{gates}/n_{test}$. Here, $n_{test}$, $TP_{inputs}$, $TP_{nodes}$ and $TP_{gates}$ represent number of tests per second, number of test patterns required to verify all inputs, nodes and gates with 100\% coverage, respectively. Note, the computations of $TP_{inputs}$, $TP_{nodes}$ and $TP_{gates}$ are presented in the Appendix~\ref{Sec:TP}.}}
	\label{fig:ge_MA_gate_node}\vspace{-15pt}
\end{figure*}

Several techniques exist for performing HT-based security vulnerability analysis, like text pattern-based analysis, model-checking-based analysis, and other formal methods-based analysis. Traditionally, these techniques analyze the equivalent behavioral, functional models, and performance models against the design constraints and functionality characteristics~\cite{chakraborty2017flexible,mohanty2017guest,rad2008sensitivity,mishra2017security,contreras2017security,ain2018formal}. However, these model-based simulations cannot guarantee complete coverage because of the computational constraints (energy and memory)~\cite{balci1997verification,pylyshyn1978computational} and floating-point inaccuracies~\cite{edwards1997design,chockler2003coverage,mancini2013system,fehnker2006formal}. 

\subsection{Motivational Analysis}
To highlight the design challenges for HT-based security vulnerability analysis, we estimated the analysis time required to perform HT-based security vulnerability analysis with 100\% coverage for all inputs, all gates, and all nodes. Note, in this HT-based security vulnerability analysis, functional and parametric behavior verification of all gates and nodes of the circuit under verification (CuV) is performed for all input patterns. By analyzing the results in Figure~\ref{fig:ge_MA_gate_node}, we made the following observations:
    
    \begin{enumerate}[leftmargin=*]
   
        \item  In HT-based security vulnerability analysis w.r.t all inputs, approximately 100x increment in the computational cost is required to reduce the analysis time (from 30 hours to 3.1 hours) of HT-based security vulnerability analysis of a small sequential circuit, i.e., 8-bit shift register (circuit B), with 100\% coverage
        \item In HT-based security vulnerability analysis w.r.t. all nodes and gates, 100x increment in the computational cost (increasing the value of $n_{test}$ 10 to 1000) does not reduce the analysis time to a realistic value. 
    \end{enumerate}
   
The above-mentioned observations lead to an important research question about \textit{how to find an efficient trade-off between computational cost and analysis time for 100\% coverage w.r.t. all inputs, all gates, and all nodes of the CuV?}

Towards this trade-off, despite the extremely-high resource demand, mathematical modeling and formal verification \cite{zhang2011case,feiten2015formal,rathmair2013hardware,lodhi2015formal,nahiyan2016avfsm,hasan2016translating,veeranna2016hardware,imran2016,abbassi2019using,abbassi2018mcsevic,pasareanu2016multi,ngo2015hardware,drechsler2004advanced} like SAT solvers and model checking based approaches provide the completeness and accuracy to some extent. However, these techniques deploy symbolic execution or fixed approximation of side-channel parameters. Therefore, they are unable to incorporate the parametric behavior, especially leakage power, and effects of process variations, which limits their scope to Active Hardware Trojans (AHTs\footnote{Note, all the abbreviations are defined in Table~\ref{Abbreviations} of Appendix~\ref{sec:LoA}}, that remain active even under the normal operation).

\subsection{Associated Research Challenges:} 
The aforementioned limitations of simulation-based, model-based, and formal verification-based techniques introduce the following research challenges:
\begin{enumerate}[leftmargin=*]
	\item \textbf{Completeness:} How to \textit{ensure 100\% coverage} in HT-based security vulnerability analysis for all gates and nodes while considering all the possible input patterns?
	\item \textbf{Feasibility:} How to \textit{reduce the analysis time} while preserving the 100\% coverage for all gates and nodes with all possible input patterns? 
	\item \textbf{Accuracy:} How to incorporate the \textit{parametric behavior and effects of process variations} to ensure an accurate HT-based security vulnerability analysis, even in the presence of \textit{Stealthy Hardware Trojans (SHTs, that remain inactive until they receive any activation trigger)}?
\end{enumerate}

\subsection{Our Novel Contribution}
\begin{figure}[!t]
    	\centering
    	\includegraphics[width=1\linewidth]{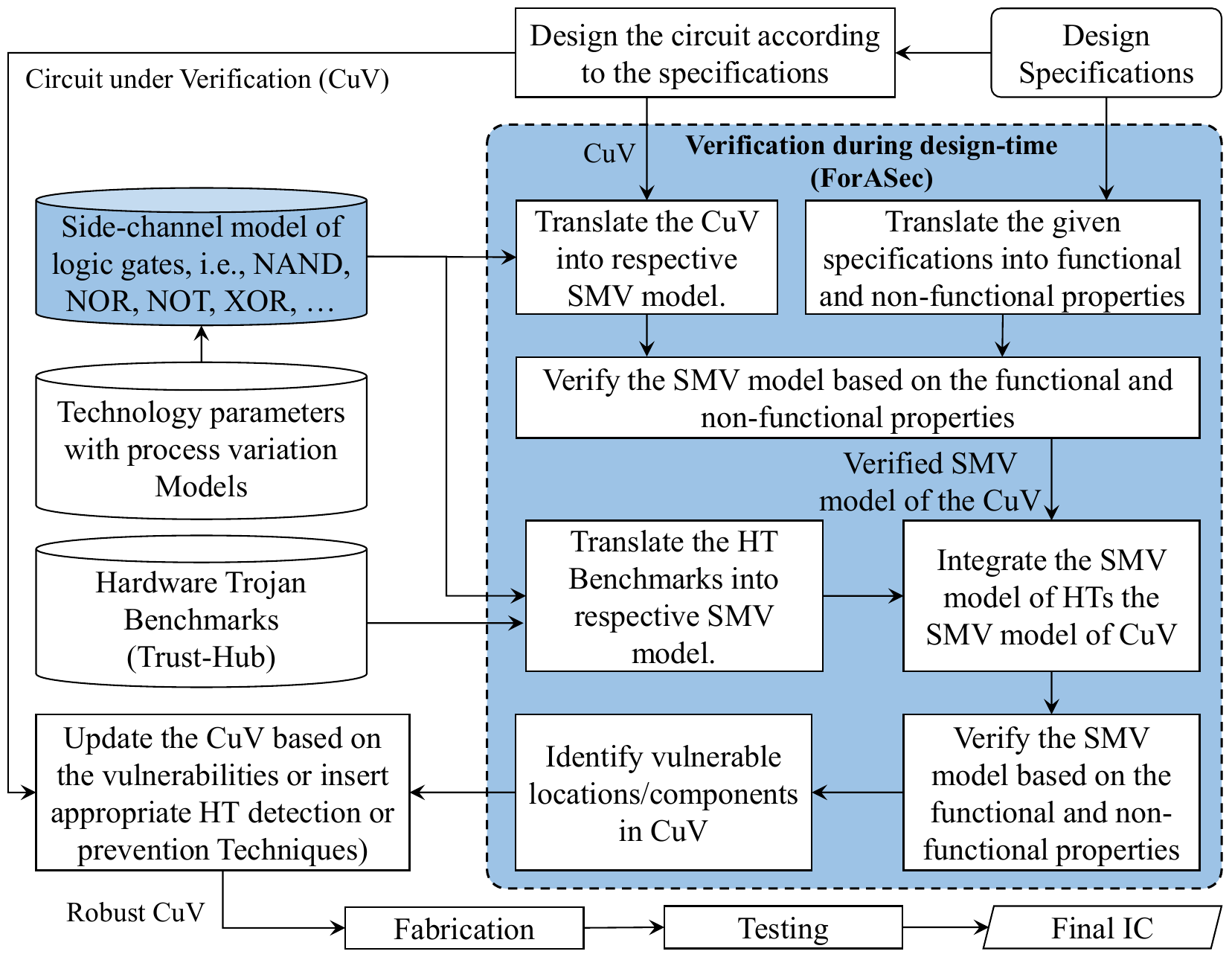} 
    	\caption{\textit{Design-time flow of our verification technique (ForAsec) for analyzing the HT-based security vulnerabilities in sequential circuits. First, it translates the CuV into the respective SMV model based on side-channel models of logic gates and translates the given specification into corresponding linear temporal logic (LTL) properties. After verification of the CuV using LT properties, it inserts the SMV models of the HT benchmarks into the verified model of the CuV. Then, it performs verification using the same LTL properties and identifies the vulnerable nodes/gates. These vulnerabilities are used to improve the CuV design. The blue-colored boxes represent our novel contributions. In this figure, SMV\protect\footnotemark stands for Symbolic Model Verification.}}
    	\label{fig:NovelContribution}
\end{figure}

To address the above challenges, we propose a novel framework for formal analysis of HT-based security vulnerabilities (ForASec). It performs a comprehensive security analysis providing the complete 100\% coverage of parametric behavior (i.e., leakage power, dynamic power and propagation delay) and process variations along with functional verification. The proposed ForASec consists of the following key features (see an overview and design flow in Fig.~\ref{fig:NovelContribution}  that clearly outlines our novel contributions and existing components of the design flow): 

\begin{table*}[]
	\centering
	\caption{\textit{Feature Comparison with State-of-the-Art (AHT: Active Hardware Trojan, SHT: Stealthy Hardware Trojan, SE: Soft Error, \\Seq: Sequential Circuits, Comb: Combinational Circuits, PV: Process Variations, SC: Side-Channel )}}
	\label{tab:Table1}
	\scriptsize
	\resizebox{1\linewidth}{!}{
		\begin{tabular}{|l|c|c|c|c|c|c|c|c|c|c|c|c|c|}
            \hline
            \rowcolor[HTML]{EFEFEF} 
            \cellcolor[HTML]{EFEFEF} & \multicolumn{3}{c|}{\cellcolor[HTML]{EFEFEF}\textbf{Methodology}} & \multicolumn{2}{c|}{\cellcolor[HTML]{EFEFEF}\textbf{Trojans}} & \cellcolor[HTML]{EFEFEF} & \multicolumn{2}{c|}{\cellcolor[HTML]{EFEFEF}\textbf{SC Parameters}} & \multicolumn{3}{c|}{\cellcolor[HTML]{EFEFEF}\textbf{Errors}} & \multicolumn{2}{c|}{\cellcolor[HTML]{EFEFEF}\textbf{Circuit Type}} \\ \cline{2-6} \cline{8-14} 
            \rowcolor[HTML]{EFEFEF} 
            \cellcolor[HTML]{EFEFEF} & \cellcolor[HTML]{EFEFEF} & \multicolumn{2}{c|}{\cellcolor[HTML]{EFEFEF}\textbf{Modeling}} & \cellcolor[HTML]{EFEFEF} & \cellcolor[HTML]{EFEFEF} & \cellcolor[HTML]{EFEFEF} & \cellcolor[HTML]{EFEFEF} & \cellcolor[HTML]{EFEFEF} & \cellcolor[HTML]{EFEFEF} & \cellcolor[HTML]{EFEFEF} & \cellcolor[HTML]{EFEFEF} & \cellcolor[HTML]{EFEFEF} & \cellcolor[HTML]{EFEFEF} \\ \cline{3-4}
            \rowcolor[HTML]{EFEFEF} 
            \multirow{-3}{*}{\cellcolor[HTML]{EFEFEF}\textbf{\begin{tabular}[c]{@{}c@{}}Related\\ Works\end{tabular}}} & \multirow{-2}{*}{\cellcolor[HTML]{EFEFEF}\textbf{Simulation}} & \textbf{Math} & \textbf{Formal} & \multirow{-2}{*}{\cellcolor[HTML]{EFEFEF}\textbf{SHT}} & \multirow{-2}{*}{\cellcolor[HTML]{EFEFEF}\textbf{AHT}} & \multirow{-3}{*}{\cellcolor[HTML]{EFEFEF}\textbf{PV}} & \multirow{-2}{*}{\cellcolor[HTML]{EFEFEF}\textbf{Delay}} & \multirow{-2}{*}{\cellcolor[HTML]{EFEFEF}\textbf{Power}} & \multirow{-2}{*}{\cellcolor[HTML]{EFEFEF}\textbf{Functional}} & \multirow{-2}{*}{\cellcolor[HTML]{EFEFEF}\textbf{Timings}} & \multirow{-2}{*}{\cellcolor[HTML]{EFEFEF}\textbf{SE}} & \multirow{-2}{*}{\cellcolor[HTML]{EFEFEF}\textbf{Seq}} & \multirow{-2}{*}{\cellcolor[HTML]{EFEFEF}\textbf{Comb}} \\ \hline
			\cite{mishra2017security,contreras2017security} & \checkmark &  &  &  & \checkmark &  & \checkmark & \checkmark & \checkmark & \checkmark &  & \checkmark & \checkmark \\ \hline
			\cite{zhang2011case} & \checkmark & \checkmark &  &  & \checkmark &  &  & & \checkmark & & & \checkmark & \checkmark \\ \hline
			\cite{veeranna2016hardware} & \checkmark &  & \checkmark &  & \checkmark &  &  & & \checkmark & & & \checkmark & \checkmark \\ \hline
			\cite{ain2018formal} &  &  & \checkmark &  &  &  & \checkmark &  & \checkmark & \checkmark & & \checkmark & \\ \hline
			\cite{feiten2015formal,rathmair2013hardware,lodhi2015formal,nahiyan2016avfsm} &  &  & \checkmark &  & \checkmark &  &  & & \checkmark & & \checkmark & \checkmark & \checkmark\\ \hline
			\cite{hasan2016translating,guo2016scalable,hasan2018novel,cruz2018hardware} &  &  & \checkmark &  & \checkmark &  &  & & \checkmark & & \checkmark & \checkmark & \checkmark\\ \hline
			\cite{ngo2015hardware,pasareanu2016multi,bobda2019synthesis,qin2018property} &  &  & \checkmark &  & \checkmark &  & \checkmark & \checkmark & \checkmark & \checkmark & & \checkmark & \checkmark \\ \hline
			\cite{imran2016,abbassi2019using} &  & \checkmark & \checkmark &  & \checkmark &  & \checkmark & \checkmark & \checkmark & \checkmark & & & \checkmark \\ \hline
			\cite{abbassi2018mcsevic} &  & \checkmark & \checkmark & \checkmark & \checkmark & \checkmark & \checkmark  & \checkmark & \checkmark & \checkmark & & & \checkmark \\ \hline
			{\cellcolor[HTML]{FFFFC7}\textbf{ForASec}} & {\cellcolor[HTML]{FFFFC7} }  & {\cellcolor[HTML]{FFFFC7}\pmb{\checkmark}} & {\cellcolor[HTML]{FFFFC7}\pmb{\checkmark}} & {\cellcolor[HTML]{FFFFC7}\pmb{\checkmark}} & {\cellcolor[HTML]{FFFFC7}\pmb{\checkmark}} & {\cellcolor[HTML]{FFFFC7}\pmb{\checkmark}} & {\cellcolor[HTML]{FFFFC7}\pmb{\checkmark}} & {\cellcolor[HTML]{FFFFC7}\pmb{\checkmark}} & {\cellcolor[HTML]{FFFFC7}\pmb{\checkmark}} & {\cellcolor[HTML]{FFFFC7}\pmb{\checkmark}} & {\cellcolor[HTML]{FFFFC7}\pmb{\checkmark}} & {\cellcolor[HTML]{FFFFC7}\pmb{\checkmark}} & {\cellcolor[HTML]{FFFFC7}\pmb{\checkmark}} \\ \hline
	\end{tabular}}\vspace{-15pt}
\end{table*}
\footnotetext{Symbolic model checking: Instead of enumerating reachable states one at a time, in state-space, Boolean encoding for state machine and sets of states are used to handle the larger and complex systems, e.g.,  binary decision diagrams (BDD). Note, SMV is the modeling language used in symbolic model checking~\cite{mcmillan1999smv}.} 
\begin{enumerate}[leftmargin=*]

    \item \textbf{Side-channel parametric modeling of the circuit under verification (CuV):} Traditional functional models can only be used to identify the vulnerabilities that are associated with a change in functionality. However, HTs have a direct or indirect impact on the side-channel behavior of ICs. Therefore, to encompass this behavior of HTs, we propose the comprehensive mathematical models of basic logic gates, i.e., NOR, NAND, and NOT, based on the behavior of side-channel parameters. Moreover, to incorporate the uncertain behavior of real-world scenarios, we also consider the process variations effects on different technology parameters, e.g., switch on-resistance ($R_{on}$), oxide capacitance ($t_{ox}$), carrier mobilities ($\mu_{n}$ and $\mu_{p}$), gate ($C_{g}$) capacitance, drain ($C_{d}$) capacitance and source ($C_{s}$) capacitance. These side-channel parameter-based gate models are used to develop the SC-based CuV model.   
        
    \item \textbf{HT-based Security vulnerability analysis}: A model checking-based methodology that uses the HT benchmarks for performing the HT-based security vulnerability analysis to identify the vulnerable nodes/gates/block. 
        
    \item \textbf{Verification/analysis algorithm for 100\% coverage of security analysis:} It is very challenging to ensure the complete coverage in security vulnerability analysis of large-sized circuits with respect to all gates, nodes, and inputs against multiple intrusions at different locations because it leads to a state-space explosion problem and increases the analysis time. Moreover, the uncertain behavior of side-channel parameters further worsens these issues. Therefore, we proposed an algorithm that consists of two phases; (a) segmentation depending upon the available RAM and (b) an HT-based security vulnerability analysis that performs the individual verification for side-channel parameter-based models using inverse LTL properties. Segmentation reduces the analysis time, but for complex circuits, the number of segments becomes very large. Therefore, to further reduce the complexity and analysis time, we propose to analyze the negation of LTL properties because the main goal of this analysis is to identify the vulnerabilities instead of correct behavior. The analysis of negated LTL properties generates counterexamples 10x faster than the analysis of non-negated LTL properties.     
\end{enumerate}

To evaluate the effectiveness of ForASec, we implement certain key basic sequential circuits and all the ISCAS89 benchmark in the presence of trust-hub Trojan benchmarks~\cite{salmani2013design,shakya2017benchmarking} for different ASIC technologies, i.e., 65nm, 45nm, and 22nm. The experimental results demonstrate that \textit{ForASec is able to correctly identify the most vulnerable node(s) and \textit{the minimum-possible size of SHTs that can be detected} while analyzing the leakage power.} Moreover, it also provides approximately 6x to 10x speedup in analysis time compared to state-of-the-art model checking-based techniques~\cite{imran2016,abbassi2019using,abbassi2018mcsevic} (See Sections~\ref{performance},~\ref{sec:MAT} and~\ref{sec:SoA}).

\section{Related Work} 
Typically, \textit{model based simulations} are used for analyzing the HT-based security vulnerabilities \cite{mishra2017security,contreras2017security}, but they cannot cover all possible test cases in complex systems because of their computational constraints (energy and memory) \cite{balci1997verification} and floating point inaccuracies \cite{edwards1997design}. To ensure the completeness and accuracy, \textit{mathematical modeling} and \textit{formal verification based vulnerability analysis} techniques have been proposed \cite{zhang2011case,feiten2015formal,rathmair2013hardware,lodhi2015formal,nahiyan2016avfsm,hasan2016translating,veeranna2016hardware,imran2016,pasareanu2016multi,ngo2015hardware}, as shown in Table \ref{tab:Table1}. Although, to some extent, mathematical modeling can overcome above-mentioned limitations, it is \textbf{still prone to human error} and \textbf{increases the design time}. On the other hand, \textit{formal verification based approaches}, i.e., SAT solving and model checking, can overcome the above stated limitations of simulation-based techniques by virtue of their inherent soundness and completeness \cite{drechsler2004advanced}. 

\textbf{The SAT solver-based approaches} are used for multistage assertion-based verification, code coverage analysis, redundant circuit removal for isolation of suspicious signals, and automatic test pattern generations in vulnerability analysis \cite{zhang2011case,feiten2015formal}. However, they provide information about the satisfaction of a particular property, but in case of a failure, they are unable to identify the reason, thus lack the debugging feature in the vulnerability analysis \cite{bao2017accelerating}.

Several \textbf{model checking-based approaches} use the functional~\cite{rathmair2013hardware,lodhi2015formal,nahiyan2016avfsm,fey2011effective,alsaiari2019hardware} and behavioral~\cite{hasan2016translating,guo2016scalable,hasan2018novel,cruz2018hardware} models of the CuV for comprehensive vulnerability analysis, and cover all possible input test cases, but \textit{model checking-based techniques inherently pose the state-space explosion problem for complex systems}. To address this issue, traditional state-space explosion reduction techniques are being used~\cite{pelanek2006reduction,dams2018abstraction}. One of the most commonly used techniques is abstraction~\cite{ho2017property}. Abstraction, in the context of model checking, can be done by focusing only on the concrete transition ($\alpha$-search algorithm), removing the dead variables (exact abstraction), or lossy compression (non-exact abstraction). Abstraction can significantly reduce the size of state-space, but in the context of security, it has the following limitations:
   
    \begin{enumerate}[leftmargin=*]
        \item By removing the dead variables from the model, it ignores the potential of using these variables as HT triggers. 
        \item Focusing only on the concrete transitions can leave a key aspect of security, i.e., exploiting the dangling transitions to perform an attack.
        \item Traditionally, abstraction ignores the rarely active parts. However, most of the HT triggers are based on the rarely active nodes. 
    \end{enumerate}
\begin{figure*}[!t]
	\centering
	\includegraphics[width=1\linewidth]{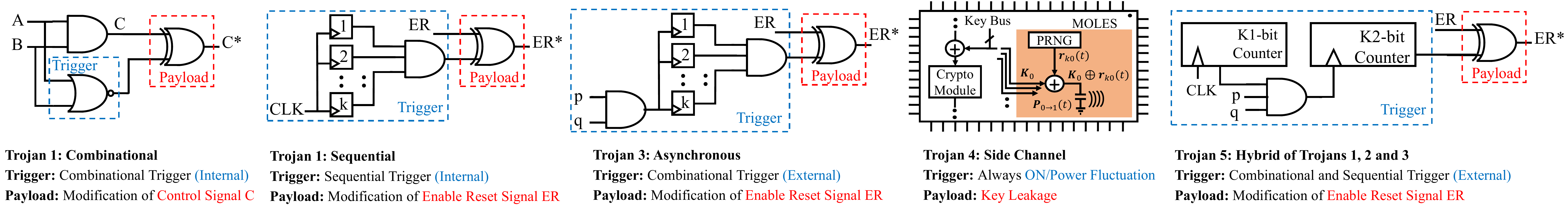} 
	\caption{Triggering and payload mechanisms of HTs}
	\label{fig:HT}\vspace{-10pt}
\end{figure*}
Another commonly used technique to address the state-space explosion is removing weak equivalence, i.e., merging the transitions or perform approximations. Similar to the abstraction, this technique also ignores the rarely active regions. Moreover, approximated modeling is vulnerable to security threats because it ignores dangling states and dead variables. Hence, these techniques cannot be directly applied for HT-based security vulnerability analysis.

To overcome this issue, several approaches for \textit{performance modeling} have been proposed, which model the side-channel parameters, i.e., power~\cite{veeranna2016hardware,imran2016,pasareanu2016multi}, propagation delay~\cite{imran2016,pasareanu2016multi} and temperature~\cite{ngo2015hardware}, and analyze the vulnerabilities based on temporal properties~\cite {imran2016,pasareanu2016multi,ngo2015hardware,bobda2019synthesis,qin2018property} and property specification language \cite{veeranna2016hardware}. These approaches provide the symbolic execution or fixed approximation of side-channel parameters. Therefore, they only target the Active Hardware Trojan-based security vulnerabilities and may not be able to analyze the security vulnerabilities due to Stealthy Hardware Trojans (SHT). These SHTs remain undetected and inactive during the testing phase, and even after deployment using external, internal, or time-based triggers, they can initiate any of the drastic payloads. The symbolic side-channel parameters of these approaches make the translated model deterministic, thus overshadows the uncertain behavior due to process variations. Moreover, most of these techniques are focused on the combinational part of the system. However, sequential circuits are also vulnerable to several security threats. To address the SHT issue, we propose a novel framework for the formal analysis of sequential circuits to identify HT-based security vulnerabilities (ForASec). It performs a comprehensive security analysis providing 100\% coverage of parametric behavior (i.e., leakage power, dynamic power, and propagation delay) and process variations.

\section{Preliminaries}
\label{sec_2}
This section provides a brief overview of different triggering mechanisms and payload of HTs, gate-level modeling, and model checking.

\subsection{Triggers and Payloads of HTs}
Fig. \ref{fig:HT} shows different types of triggering and payload mechanisms for hardware Trojans, e.g., the payload of HT 1, 2, 4, and 5 are used to hack the signals C and ER, which can be any control signal having the combinational, sequential (counter), asynchronous (counter) and hybrid triggering mechanism, respectively. However, the payload of HT 3 is to leak the information via side-channel parameters, which in this case is dynamic or leakage power. Therefore, HTs can broadly be categorized as (1) \textit{Functional HTs}, which change the system functionality by addition or deletion of modules with malicious intent, and (2) \textit{Parametric HTs}, which reduce reliability, increase the likelihood of system failure, and modify the physical parameters, such as power consumption and resulting in faster aging than expected.

\subsection{Model Checking and nuXmv}
Model-checking \cite{hasan2015formal} is primarily used as a verification technique for reactive systems by translating them into a corresponding state-space model and temporal properties. However, it provides an automatic and exhaustive verification but\textit{in complex systems, the state-space grows exponentially,} which makes it computationally impossible to explore the entire state-space with limited resources. This problem, termed as \textbf{state-space explosion}, is usually resolved by using efficient algorithms and bounded model checking (BMC) \cite{hasan2015formal}. In this paper, we use an open-source symbolic model checker, nuXmv \cite{nuxmv2014}, which supports rational number analysis and property language specification using computation tree and linear temporal logic to facilitate the modeling and verification of designs that exhibit continuous behaviors. 

\subsection{Gate-Level Performance Parameters}
In this work, we employ the following side-channel parameters for gate level modeling of complex sequential circuits~\cite{ghosh1992estimation}:

\textit{Dynamic Power} is the power dissipated while charging/discharging of load capacitances associated with transistors, nodes and wires and usually modeled as:
\begin{equation}
    \baselineskip=0pt
    \scriptsize
    \mathit{P_{switching} = \alpha.C_{total}.{V_{dd}}^{2}.f}
    \label{equation:1}
\end{equation}
Where $\mathit{\alpha}$, $\mathit{f}$, $\mathit{V_{dd}}$ and $\mathit{C_{total}}$ are the switching activity, operating clock frequency, supply voltage, and the total capacitance that is (dis)charged in a transition, respectively.

\textit{Subthreshold Leakage Power} is dissipated in an IC primarily due to the undesirable flow of subthreshold current in the channel from $\mathit{V_{dd}}$ to ground nodes, when transistors are in \emph{Off} state. In this paper, we modeled it as 
\begin{equation}
    \baselineskip=0pt
    \scriptsize
    \mathit{P_{Leakage} = V_{dd}.I_{Leakage}}.
    \label{equation:3}
\end{equation}
Equation \ref{equation:4} shows the sub-threshold leakage current of a MOSFET per transistor width, where $\mathit{W}$ is the gate width, $\mathit{L}$ is the effective channel length, $\mathit{n}$ is subthreshold slope factor, $\mathit{C_{ox}}$ is oxide capacitance, $\mathit{\phi{_t} = KT/q}$ is thermal voltage, $\mathit{\mu}$ is effective carrier mobility, and $\mathit{\sigma}$ is drain induced barrier lowering (DIBL) factor.       
\begin{equation}
    \baselineskip=0pt
    \scriptsize
    \mathit{I_{Leakage} = 2.n.\mu.C_{ox}.\frac{W}{L}.{\phi{_t}}^2.exp(\frac{\sigma.V_{dd}-V_{th}}{n.\phi{_t}})}
    \label{equation:4}
\end{equation}
\begin{figure*}[!t]
	\centering
	\includegraphics[width=1\textwidth]{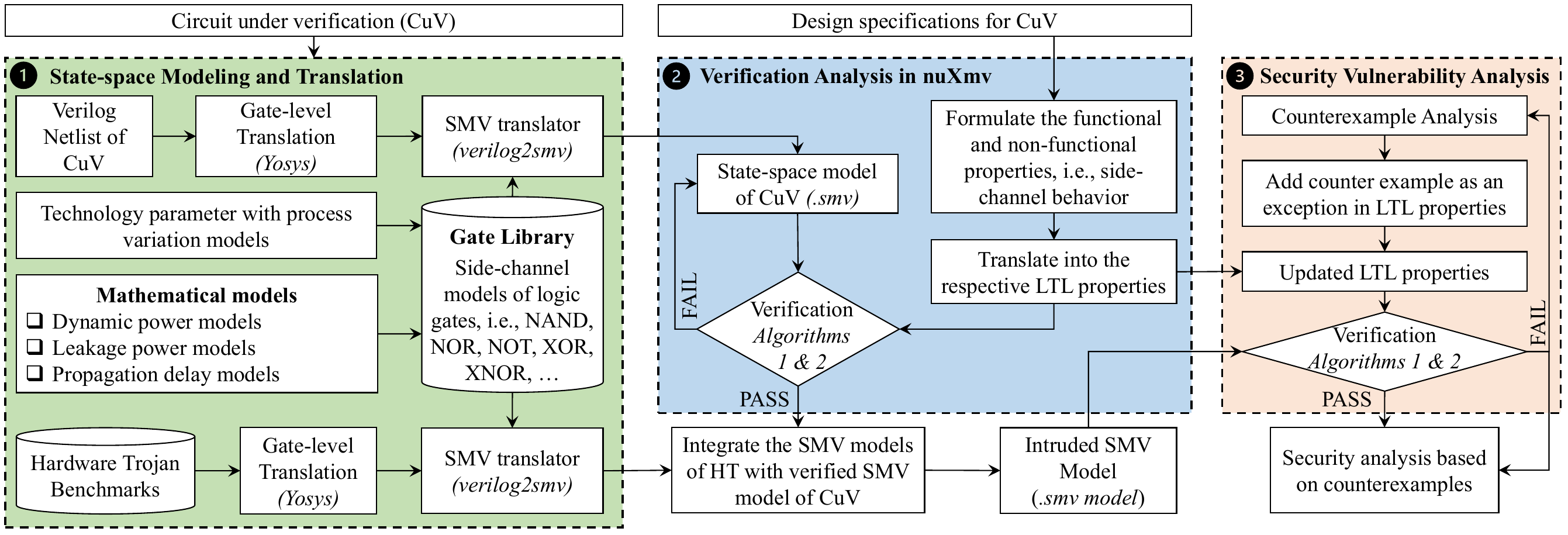} 
	\caption{\textit{The proposed ForASec Framework to perform the security analysis while considering the complete coverage of parametric behavior (i.e., leakage power, dynamic power and propagation delay) and process variations.}}
	\label{figure:methodology}
\end{figure*}
\textit{Propagation Delay} is defined as the (dis)charging time of the internal parasitic and external load capacitances. It is estimated using  \textit{Elmore delay} based on individual input transitions. 
\begin{equation}
    \baselineskip=0pt
    \scriptsize
    \mathit{t_{delay} = \ln2.\tau_{elmore}}
    \label{equation:4b}
\end{equation}
\begin{equation}
    \baselineskip=0pt
    \scriptsize
    \mathit{\tau_{elmore} = \sum_{k = 1}^{N}R_{ik}.C_k}
    \label{equation:4c}
\end{equation}
Where $\mathit{C_k}$ is the capacitance at node $\mathit{k}$, and $\mathit{R_{ik}}$ is shared resistance among the paths from root to node $\mathit{k}$ and leaf $\mathit{i}$.

\section{Threat Model}
The threat model is based on an un-trusted fab, i.e., the foundry is un-trusted while the rest of the agents (i.e., IP provider, designer, testing/verification engineer) in the IC design cycle are trusted. Such a threat model is very common in the HT community and has been widely used by prominent groups like~\cite{xiao2016hardware,cruz2018hardware}. The attacker has access to the netlist of the design. On the other hand, the main goal of the defender (IC designer) is to identify the vulnerable gates/nodes/ blocks in the design to make it robust against potential HT-based security vulnerabilities. It is important to note that the proposed technique does not detect the hardware Trojan during the design-time or testing time.
\section{ForASec: Analysis Framework }\label{sec_forasec}
ForASec analyzes the HT-based security vulnerabilities in a sequential circuit by translating it into its corresponding state-space model. Fig. \ref{figure:methodology} shows the different phases of vulnerability assessment in the proposed framework, which are explained below in Sections \ref{phase1} and \ref{phase2}. 
\subsection{State-Space Modeling and Translation} \label{phase1}
\begin{figure}[!t]
\baselineskip=0pt
 	\centering
 	\includegraphics[width=1\linewidth]{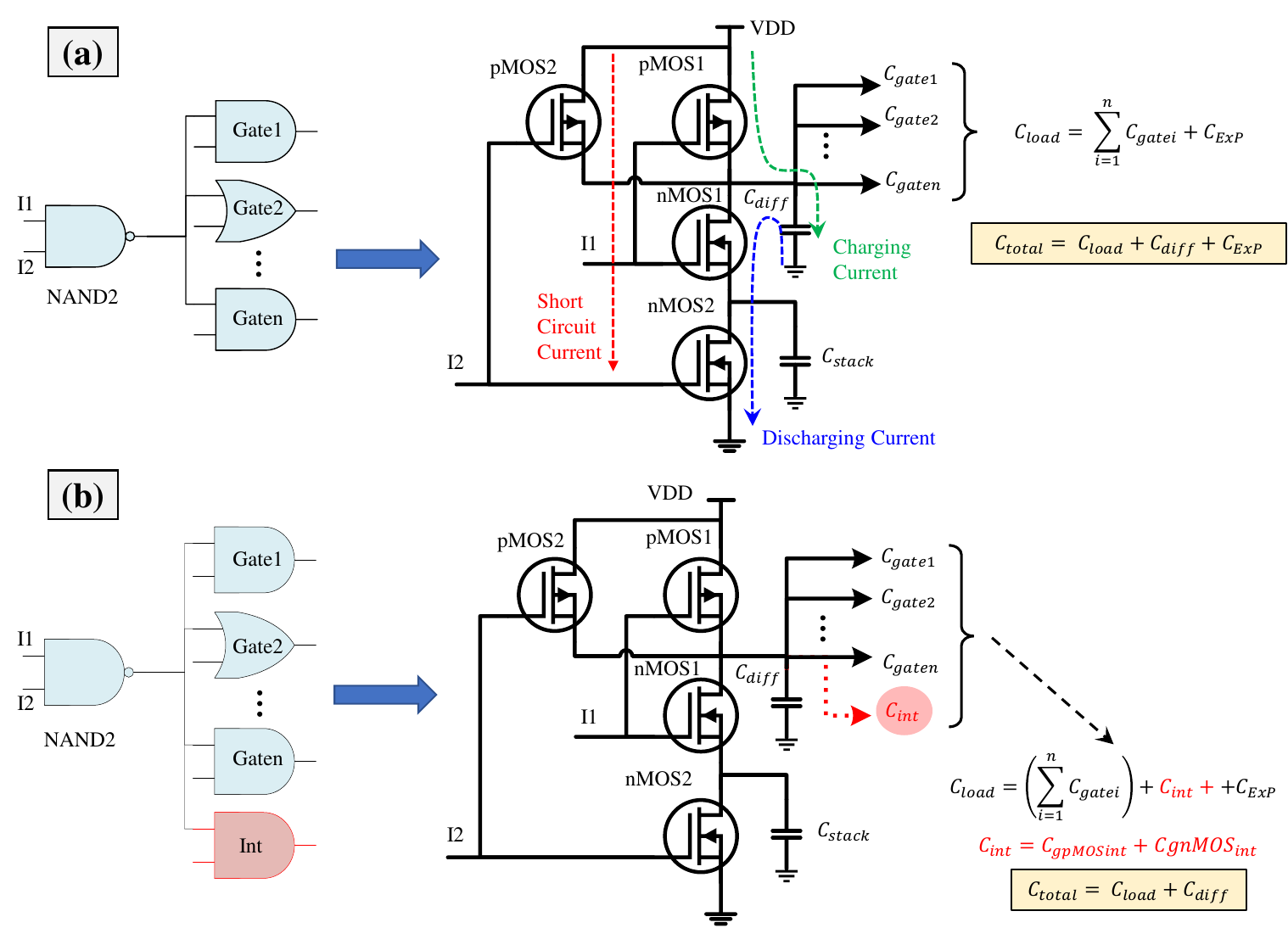}
 	\caption{\textit{Modeling of 2-Input NAND Gate without (a) and with (b) intrusions.}}
 	\label{fig:NAND}
\end{figure}

In the first phase, ForASec translates the circuits under test into their corresponding state-space and behavioral/functional/performance properties (see Step 1 in Fig. \ref{figure:methodology}). For this translation, we developed the side-channel and technology parameters-based models for all the universal 2-input gates, i.e., NAND, NOR, and NOT, that are used to model the multi-input complex gates and modules. For illustration purposes, we describe the two-input NAND gate model with different parameter values based on their transistor-level structure, shown in Fig. \ref{fig:NAND}(b).
\begin{table*}[!tbh]
	\caption{\textit{Leakage Power modeling of a 2-Input NAND Gate}}
	\label{tab:LP_NAND}
	\centering
	\footnotesize
	\resizebox{0.9\textwidth}{!}{
		\begin{tabular}{|c|c|} 
			\hline
			\textbf{State} & \textbf{Subthreshold Leakage Power} \\
			\hline
			00   &  $2.FO.V_{dd}.\big(n_n.\mu_n.C_{ox}.WR_n.\frac{Wnmin}{Ln}.{\phi{_t}}^2.exp(\frac{\sigma_n.V_{dd}-V_{thn}}{n_n.\phi{_t}})\big).10^{\frac{-(V_{dd}.\sigma_n)}{n_n}} $       \\
			\hline
			01   &  $2.FO.V_{dd}.\big(n_n.\mu_n.C_{ox}.WR_n.\frac{Wnmin}{Ln}.{\phi{_t}}^2.exp(\frac{\sigma_n.V_{dd}-V_{thn}}{n_n.\phi{_t}}) + n_p.\mu_p.C_{ox}.WR_p.\frac{Wpmin}{Lp}.{\phi{_t}}^2.exp(\frac{\sigma_p.V_{dd}-V_{thp}}{n_p.\phi{_t}})\big)$ \\
			\hline
			10   &  $2.FO.V_{dd}.\big(n_n.\mu_n.C_{ox}.WR_n.\frac{Wnmin}{Ln}.{\phi{_t}}^2.exp(\frac{\sigma_n.V_{dd}-V_{thn}}{n_n.\phi{_t}}) + n_p.\mu_p.C_{ox}.WR_p.\frac{Wpmin}{Lp}.{\phi{_t}}^2.exp(\frac{\sigma_p.V_{dd}-V_{thp}}{n_p.\phi{_t}})\big)$   \\
			\hline
			11   &  $4.FO.V_{dd}.\big(n_p.\mu_p.C_{ox}.WR_p.\frac{Wpmin}{Lp}.{\phi{_t}}^2.exp(\frac{\sigma_p.V_{dd}-V_{thp}}{n_p.\phi{_t}})\big)$   \\
			\hline
	\end{tabular}}
\end{table*}
\textbf{Dynamic power }for 2-input NAND is modeled as {\footnotesize $\mathit{P_{switch} = \alpha.C_{total}.{V_{dd}}^{2}.f}$}, where switching activity ($\alpha$), is computed based on the input/output transition probability and the total capacitance at the output of NAND is equal to $\mathit{C_{total} = C_{load} + C_{diff}}$. 

The \textit{load capacitance} ($\mathit{C_{load}}$) at the output of a single gate is the sum of gate capacitances of individual gates connected at the output node and external parasitic capacitances ($C_{ExP}$), and it is modeled as

\begin{equation}
\baselineskip=0pt
    \scriptsize
    \mathit{C_{load} = \sum\limits_{i=1}^p C_{gpMOSi} + \sum\limits_{j=1}^n  C_{gnMOSj}} + C_{ExP} 
    \label{equation:5a}
\end{equation}

The gate capacitance for an individual pMOS/nMOS transistor is modeled as 

\begin{equation}
\baselineskip=0pt
    \scriptsize
    \mathit{C_{gMOS} = FO.WR.(C_{GSO} + C_{GDO}+ W_{min}.L.C_{ox})}
    \label{equation:5b}
\end{equation}
Where $\mathit{FO}$, $\mathit{WR}$, $\mathit{C_{GDO}}$, $\mathit{C_{GSO}}$, $\mathit{W_{min}}$, $\mathit{L}$ and $\mathit{C_{ox}}$ represent fanout, width ratio, overlap capacitances, minimum width, effective length, and oxide capacitance per unit area of the gate, respectively.

To compute the external parasitic capacitance ($C_{Exp}$), we use the following standard formula to estimate the capacitance of the interconnect structures~\cite{yuan1982simple}:

\begin{equation}
\baselineskip=0pt
\scriptsize
    C_{ExP} = \epsilon \Bigg[\frac{w-0.5t}{h} + \frac{2\pi}{ln{\bigg( 1+ \frac{2h}{t} +\sqrt{\frac{2h}{t}(\frac{2h}{t} + 2)}\bigg )}} \Bigg]\ for\ w\geq \frac{t}{2}
    \label{equation:5c}
\end{equation}

\begin{equation}
\baselineskip=0pt
\scriptsize
    C_{ExP} = \epsilon \Bigg[\frac{w}{h} + \frac{\pi(1-0.0543\times\frac{t}{2h})}{ln{\bigg( 1+ \frac{2h}{t} +\sqrt{\frac{2h}{t}(\frac{2h}{t} + 2)}\bigg )}} + 1.47\Bigg]\ for\ w<\frac{t}{2}
    \label{equation:5d}
\end{equation}

Where $w$ and $t$ represent the width and thickness of the wire and $h$ represents the gap between two wires or distance from the substrate. Note, in this work, we used Equation~\ref{equation:5c} because due to design rule checks of real-world ICs, it is very uncommon that $w<\frac{t}{2}$.   

The \textit{internal diffusion capacitance} ($C_{diff}$) depends on both the sidewall perimeter $\mathit{PS}$ and area $\mathit{AS}$ of the source (drain) diffusion region and modeled as  
\begin{equation}
\baselineskip=0pt
\scriptsize
\begin{aligned}
    \mathit{ C_{diff} =~} & \mathit{n.\big(2.FO.WR_{p}.W_{pmin}.C_{pdmin} \big) } \\
    & \mathit{+ p. \big( FO.WR_{n}.W_{nmin}.C_{ndmin} \big) } 
\end{aligned}
\label{equation:5}
\end{equation}
where $\mathit{C_{pdmin}}$, $\mathit{C_{ndmin}}$, $n$ and $p$ represent the minimum intrinsic capacitance for the pMOS and nMOS transistors, the number of nMos and PMOs at the output node of the gate, respectively. The minimum drain diffusion capacitance ($\mathit{C_{dmin}}$) for a single transistor is calculated as
\begin{equation}
    \baselineskip=0pt
    \scriptsize
    \mathit{C_{dmin} = AS.C_{jbd} + PS.C_{jbsdw}} 
    \label{equation:7}
\end{equation}
where $\mathit{C_{jbd}}$ is the capacitance per unit area between body and bottom of the drain, and $\mathit{C_{jbsdw}}$ is the capacitance per unit length of the junction between body and sidewalls of the drain. Similarly, the minimum source diffusion capacitance ($\mathit{C_{smin}}$) can be computed as well. 

We have modeled the \textbf{propagation delay} as \textit{Elmore's delay} estimation because it considers the linear behavior of transistors to incorporate the effects of process variations. It is defined as the time required for charging and discharging the total capacitances via the RC tree path from drain to source. This resistive behavior of the transistors is modeled as 
\begin{equation}
    \baselineskip=0pt
    \scriptsize
    \mathit{R_{on} = \frac{L}{\mu.C_{ox}.W.(V_{gs}-V_{th})}}
    \label{equation:6}
\end{equation}
We have modeled the resistances for individual pMOS and nMOS transistors in accordance with the configuration of universal gates on different inputs. Table \ref{tab:PD_NAND} indicates the Elmore delay relation for 6 possible transitions.

We have modeled the \textbf{leakage current} at the gate level using the components mentioned in Equation \ref{equation:4}. The total leakage in an IC is the sum of leakage power of individual nMOS and pMOS transistors that are in the \emph{Off} state, which mainly depends on the input vector. In CMOS-based ICs, half of the total transistors are always in the \emph{Off} state for any given vector. The stacking of two \emph{Off} transistors in series significantly reduces the sub-threshold leakage compared to a single \emph{Off} transistor. The two nMOS transistors in the NAND gate, shown in Fig. \ref{fig:NAND}, are in series, and as a result of the stack effect on input vector 00, the total leakage current reduces by a factor 10$\mathit{^{\frac{-(V_{dd}.\sigma)}{n}}}$ by virtue of Kirchhoff's current law. The leakage power for the two input NAND gates can be modeled with respect to the input states as shown in Table \ref{tab:LP_NAND}. 

In case of \textit{intrusion} at the output of any gate, the total capacitance $(C_{total})$ of the gate can increase with the factor $C_{int}$, which is the gate capacitance of the intruded gates ($\mathit{C_{int} = C_{int1} + C_{int2} + ... + C_{intn}}$). Thus, the total capacitance $\mathit{C_{total}}$ at the output of the gate is changed to $\mathit{C_{total} = C_{load} + C_{diffusion} + C_{int} + C_{ExP}}$. This increase in capacitance can affect the overall performance of the gate in terms of side-channel parameters. For example, if the intruded gate is within the propagation path then it has a major effect on the propagation delay and dynamic power, otherwise its effect is more on switching and leakage power compared to propagation delay.
\begin{table}[!t]
	\small
	\renewcommand{\arraystretch}{1.5}
	\caption{\textit{Propagation Delay Modeling of 2-Input NAND Gate}}
	\label{tab:PD_NAND}
	\centering
	\footnotesize
	\resizebox{0.9\linewidth}{!}{
		\begin{tabular}{|c|c|l|}
			\hline
			\begin{tabular}[c]{@{}l@{}}I/P Transition\end{tabular} & \begin{tabular}[c]{@{}l@{}}O/P Transition\end{tabular} & Elmore Delay ($\tau$) \\ \hline
			01 $\rightarrow$ 11  & 1 $\rightarrow$ 0 & $2.R_n.C_{total} / FO.WR_n$ \\ \hline
			10 $\rightarrow$ 11  & 1 $\rightarrow$ 0 & $2.R_n.(C_{total} + C_{nstack}) / FO.WR_n$ \\ \hline
			00 $\rightarrow$ 11  & 1 $\rightarrow$ 0 & $2.R_n.C_{total} / FO.WR_n$         \\ \hline
			11 $\rightarrow$ 01  & 0 $\rightarrow$ 1 & $R_p.C_{total} / FO.WR_p$          \\ \hline
			11 $\rightarrow$ 00  & 0 $\rightarrow$ 1 & $R_p.C_{total} / 2.FO.WR_p$         \\ \hline
			11 $\rightarrow$ 10  & 0 $\rightarrow$ 1 & $R_p.(C_{total} + C_{nstack}) / FO.WR_p$          \\ \hline
	\end{tabular}}
\end{table} 
\subsection{Verification and Vulnerability Analysis}\label{phase2}
In the second phase, ForASec verifies the translated model and intrudes it to analyze the vulnerable locations (see Fig. \ref{figure:methodology}). First, the functional, behavioral, and performance tests are performed to ensure the correctness of the translated model (see Step 2 in Fig. \ref{figure:methodology}). We assume that an attacker has access to the netlist. Therefore, to model this behavior, \textit{a different number of intruded gates are applied at different locations in the model} to \textbf{ generate counterexamples} for the given functionality, behavior, and performance. These counterexamples are used to analyze and identify the vulnerabilities for undesired behavior caused due to malicious alterations (see Step 3 in Fig. \ref{figure:methodology}).

We propose a generic set of functional, behavioral, and performance properties based on the CuV, to analyze the vulnerabilities against intrusions based on the performance bounds of the circuits, which are calculated by applying different distributions of technology parameters. Note: the lower bounds for dynamic power and path delays are not considered because these values are zero in the steady-state condition. The above-mentioned bounds are computed under the following definitions:       
\begin{enumerate}[leftmargin=*]
	\item The upper bound of dynamic power is modeled as  $\mathit{\sum_{i=1}^{n} DP_{imax}}$, where $\mathit{DP_{imax}}$ is the maximum dynamic power of an $i^{th}$ gate.
	\item The upper bound of the delay of the $\mathit{k^{th}}$ path is modeled as $\mathit{ \sum_{i=1}^{m} Delay_{imax}}$, where $\mathit{Delay_{imax}}$ is the maximum delay of the $i^{th}$ gate in the selected path. 
	\item The upper/lower bound of leakage power is modeled as $\mathit{\sum_{i=1}^{n} LP_{imax/imin}}$, where $\mathit{LP_{imax/imin}}$ is the maximum/ minimum leakage power dissipated by an $i^{th}$ gate.
\end{enumerate}
\subsection{Algorithm for HT-based Security Vulnerability Analysis} 
It analyzes the state-space model of the given circuit and generates the corresponding counterexamples using the set of linear temporal logical (LTL) properties. These properties are based on the upper and lower bound of side-channel parameters, like propagation delay, dynamic, and leakage power. To capture the effect on \textit{dynamic power}, we propose property \ref{property:1} which states that there exists a state in which dynamic power is out of upper bound, which can be true if any intrusion has a significant impact on dynamic power. However, if the property is false, then it indicates the scenario in which either there is no intrusion in the IC or the dynamic power fails to detect the HTs.
\begin{algorithm}[!t]
    \footnotesize
    	\caption{\footnotesize{\textbf{:} HT-based Security Vulnerability Analysis }}
    	\label{algo:sec_analysis}
    	\begin{algorithmic}[1]
    		\Input
    		\Statex Maximum dynamic power ($\mathit{DP_{max}}$)
    		\Statex Maximum leakage power $\mathit{LP_{max}}$
    		\Statex Minimum leakage power ($\mathit{LP_{min}}$)
    		\Statex Maximum delays for k paths of IC ($\mathit{d1_{max}..dk_{max}}$)
    		\Statex $M_{DP}$: Dynamic power model of CuV
    		\Statex $M_{LP}$: Leakage power model of CuV
    		\Statex $M_{PD}$: Propagation delay model of CuV
    		\Statex $M_{F}$: Functional model of CuV
    		\Statex $H_{DP}$: Dynamic power model of HT benchmarks
    		\Statex $H_{LP}$: Leakage power model of HT benchmarks
    		\Statex $H_{PD}$: Propagation delay model of HT benchmarks
    		\Statex $H_{F}$: Functional model of HT benchmarks 
    		\Output 
    		\Statex Vulnerable locations
    		\State $M_{DP} = M_{DP} + H_{DP}$
    		\State $M_{LP} = M_{LP} + H_{LP}$
    		\State $M_{D} = M_{D} + H_{D}$
    		\State $M_{F} = M_{F} + H_{F}$
    		\State Perform the functional verification using $M_{F}$
    		\If{$M_{F}$ is verified}
    		\While{Property~\ref{property:1} = TRUE} // upper bound of dynamic power
    		\State Generate counterexample using Property~\ref{property:2} for $M_{DP}$;
    		\State Add counterexample as an exception;
    		\EndWhile
    		\While{Property~\ref{property:3} = TRUE} // upper bound of propagation delay
    		\State Generate counterexample using Property~\ref{property:4} for $M_{PD}$;
    		\State Add counterexample as an exception;
    		\EndWhile			
    		\While{Property~\ref{property:5} = TRUE} // upper/lower bound of leakage power
    		\State Generate counterexample using Property~\ref{property:7} for $M_{LP}$;
    		\State Add counterexample as an exception;
    		\EndWhile
    		\Else 
    		\State Update $M_{DP}$, $M_{LP}$, $M_{D}$, and $M_{F}$ 
    		\EndIf
    	\end{algorithmic}
    \end{algorithm}
\begin{align*}
	\baselineskip=0pt
	\footnotesize
	\texttt{$F(DP\;!= 0 \rightarrow DP_{MAX} < DP_1 + DP_2 +...+ DP_n)$}  
	\tag{\footnotesize I}\label{property:1}
\end{align*}
Once the unwanted behavior of intrusions is identified (Property \ref{property:1} holds), we take the complement of Property \ref{property:1}, to generate the error trace, as shown in Property \ref{property:2}. 
\begin{align*}
	\baselineskip=0pt
	\footnotesize
	\texttt{$G(DP\;!= 0 \rightarrow DP_{MAX} >= DP_1 + DP_2 +...+ DP_n)$} 
	\tag{\footnotesize II}\label{property:2}
\end{align*}

The counterexample driven analysis is performed to identify the vulnerable locations. After analyzing the counterexample, it is added as an exception in Property \ref{property:1} to generate Property \ref{property:9}. This process repeats until the Property \ref{property:1} becomes false, as shown in line 1 to 4.
\begin{align*}
	\baselineskip=10pt
	\footnotesize
	\texttt{$F((DP\;!= 0)\&(C_1\&..\& C_n)\rightarrow DP_{MAX} < DP_1 +...+ DP_n)$}  
	\tag{\footnotesize III}\label{property:9}
\end{align*}
Similarly, this iterative process is applied on the following set of LTL properties to analyze effects on the \textit{propagation delay} and \textit{leakage power}.  
\begin{align*}
	\baselineskip=10pt
	\footnotesize
	\texttt{$F(D(k) \;!= 0 \rightarrow D(k)_{max} < D_1(k) +...+ D_n(k))$}  
	\tag{\footnotesize IV}\label{property:3}
\end{align*}   
\begin{align*}
	\baselineskip=10pt
	\footnotesize
	\texttt{$G(D(k) \;!= 0 \rightarrow D(k)_{max} >= D_1(k) +...+ D_n(k))$}
	\tag{\footnotesize V}\label{property:4}
\end{align*} 
\begin{align*}
	\baselineskip=10pt
	\footnotesize
	\texttt{$F\mathit{(LP_{max/min} </> LP_1 + LP_2 +...+ LP_n)}$}  
	\tag{\footnotesize VI}\label{property:5}
\end{align*}
\begin{align*}
	\baselineskip=10pt
	\footnotesize
	\texttt{$G\mathit{(LP_{max/min} >= /<= LP_1 + LP_2 +...+ LP_n)}$}  
	\tag{\footnotesize VII}\label{property:7}
\end{align*} 

The process to identify the vulnerable nodes/gates using counterexample analysis consists of the following key steps:
\begin{enumerate}[leftmargin=*]
    \item The first step is to generate the counterexamples for CuV using the inverse of LTL properties for all parameters, dynamic power, leakage power, and propagation delay. The reason behind this step is to estimate the parametric behavior of each gate. 
    \item In the next step, the same process is repeated after intruding the CuV. In nuXmv, the generated counterexample provides the values of each parameter for every gate. Therefore, we can compare the values of these parameters with the parametric behavior of the un-intruded CuV. In the results of this comparison, we can pinpoint the vulnerable gates. Moreover, a counterexample generated using the propagation-related properties provides the values of $C_{total}$ and $C_{load}$ as well. Therefore, comparing these values with un-intruded CuV can highlight the potential nodes affected by the intrusions.   
\end{enumerate}
Note, by increasing the number of intruded gates in the above-mentioned analysis, we can determine the maximum size of the SHTs that does not generate any counterexample. Hence, we get the maximum size of SHT that can remain undetected to HT detection techniques that use dynamic power, leakage power, or propagation delay.

\subsection{State-Space Explosion} 
For complex and large circuits, the number of variables in the model along with the individual gates increases significantly, which escalates the complexity, and thus the above-mentioned counterexample analysis may take a long time. \textit{However, our proposed algorithm and the modular structure of gate models reduce the complexity by enabling to construct and analyze the dynamic power, leakage power and delay models separately.} However, this may still not be sufficient to address the state-space explosion problem. Therefore, we proposed a novel method which can be further leveraged depending on the complexity and size of the given sequential circuits. The proposed methodology consists of the following steps: 
\begin{algorithm}[!t]
    \footnotesize
    \caption{\footnotesize{\textbf{:} Model Segmentation to avoid State-Space Explosion}}
    \label{algo:SS_explosion}
    \begin{algorithmic}[1]
    \Input
    \Statex $M_{DP}$: Dynamic power model of CuV
    \Statex $M_{LP}$: Leakage power model of CuV
    \Statex $M_{PD}$: Propagation delay model of CuV
    \Statex $M_{F}$: Functional model of CuV
    \Statex $RAM$: Available RAM
    \Output 
    \Statex Vulnerable locations
    \State Generate the binary decision diagram for $M_{F}$  
    \State $ME(M_{F})\ \gets\ size(M_{F}) < RAM$
    \State $i = 1$
    \While{$ME(M_{F})$ = TURE}
        \State Divide $M_{F}$ into 2 segments
        \State $M_{F} := {f_1, f_2,...,f_{2^{i}}}$
        \State Generate the binary decision diagram for each segment of $M_{F}$  
        \State $ME(M_{F})\ \gets\ max(size(M_{F})) < RAM$
        \State $i = i+1$
    \EndWhile
    		
    \State Divide $M_{DP}$ into $2^i$ segments
    \State $M_{DP} := {dp_1, dp_2,...,dp_{2^{i}}}$
    \State Verify the inversion of Property~\ref{property:1} for each segment of $M_{DP}$  
    \State $ME(M_{DP})\ \gets\ max(size(M_{DP})) < RAM$
    \State $j = 1$
    \While{$ME(M_{DP})$ = TRUE}
        \State Divide $M_{DP}$ into 2 segments
        \State $M_{DP} := {dp_1, dp_2,...,dp_{2^{i+j}}}$
        \State Verify the inversion of Property~\ref{property:1} for each segment of $M_{DP}$  
        \State $ME(M_{DP})\ \gets\ max(size(M_{DP})) < RAM$
        \State $j = j+1$
    \EndWhile
    		
    \State Divide $M_{LP}$ into $2^{i+j}$ segments
    \State $M_{LP} := {lp_1, lp_2,...,lp_{2^{i+j}}}$
    \State Verify the inversion of Property~\ref{property:5} for each segment of $M_{LP}$  
    \State $ME(M_{LP})\ \gets\ max(size(M_{LP})) < RAM$
    \State $k = 1$
    \While{$ME(M_{LP})$ = TRUE}
        \State Divide $M_{LP}$ into 2 segments
        \State $M_{LP} := {lp_1, lp_2,...,lp_{2^{i+j+k}}}$
        \State Verify the inversion of Property~\ref{property:5} for each segment of $M_{LP}$ 
        \State $ME(M_{LP})\ \gets\ max(size(M_{LP})) < RAM$
        \State $k = k+1$
    \EndWhile
    		
    		\State Divide $M_{D}$ into $2^{i+j+k}$ segments
    		\State $M_{D} := {d_1, d_2,...,d_{2^{i+j+k}}}$
    		\State Verify the inversion of Property~\ref{property:3} for each segment of $M_{D}$  
    		\State $ME(M_{D})\ \gets\ max(size(M_{MD})) < RAM$
    		\State $l = 1$
    		\While{$ME(M_{D})$ = TRUE}
    		\State Divide $M_{D}$ into 2 segments
    		\State $M_{D} := {d_1, d_2,...,d_{2^{i+j+k+l}}}$
    		\State Verify the inversion of Property~\ref{property:3} for each segment of $M_{D}$   
    		\State $ME(M_{D})\ \gets\ max(size(M_{D})) < RAM$
    		\State $l = l+1$
    		\EndWhile
    		
            \State Apply Algorithm~\ref{algo:sec_analysis} on each segment of $M_{DP}$, $M_{LP}$, $M_{D}$, and $M_{F}$
    	\end{algorithmic}
    \end{algorithm}

\textbf{Segmentation of the functional and side-channel models:} The first step of this methodology is dividing the side-channel and functional models based on the available Random Access Memory (RAM), as depicted in Algorithm~\ref{algo:SS_explosion}. The key steps of this algorithm are:
        
\begin{enumerate}[leftmargin=*]
    \item First, the switching activities ($\alpha$) of each node in the CuV are obtained from Verilog file by generating a .saif file. These switching activities model the impact of other segments on the power consumption and propagation delay. 
    \item Estimate the size of the functional model ($M_{F}$) of the CuV and compare it with the available RAM using verilog2smv translator. If the size of $M_{f}$ is larger than the available RAM, then divide the model into two segments and perform a memory check for the largest segment. Note, the segmentation process is done in Verilog, and the Verilog segments are translated into respective SMV segments after a successful functionality test. Repeat this process until there is no memory error for $M_{F}$. Moreover, to model the effects of segments on each other, the probability of the inputs of each segment is computed using the switching activities, see Table~\ref{tab_SAct}. To ensure the correctness of this translation, switching activity on each node of the SMV model is computed and compared with the .saif file. If the switching activity of any node is not equal to the corresponding value in the .saif file, then the segmentation process is repeated.  
   
    \begin{table}[!t]
        \centering
    	\caption{\textit{Switching Activities of basic gates.}}
    	\label{tab_SAct}
    	\resizebox{1\linewidth}{!}{
        \renewcommand{\arraystretch}{1.3}
        \begin{tabular}{|l|l|l|l|l|l|}
            \hline
            Gates & $P_{Y}$ & Switching Activity $\alpha_{Y}$ & Gates & $P_{Y}$ & Switching Activity $\alpha_{Y}$ \\ \hline
            2-input NAND  & $ 1 - P_{A}P_{B}$ & $(1 - P_{A}P_{B}) \times (P_{A}P_{B})$ & NOT gate  & $\bar{P_{A}}$ & $\bar{P_{A}}P_{A}$\\ 
            2-input NOR  & $\bar{P_{A}} \bar{P_{B}}$ &  $(\bar{P_{A}}\bar{P_{B}})\times (1 - \bar{P_{A}}\bar{P_{B}})$ & & & \\ \hline
        \end{tabular}}
    \end{table}
    \item After segmenting the $M_{F}$, divide the dynamic power model of the CuV ($M_{DP}$) and perform a memory check for the largest segment by verifying the inversion of Property~\ref{property:1} (see line 11-14 in Algorithm~\ref{algo:SS_explosion}). If there is a memory error, then divide the model into two segments and perform a memory check for the largest segment. Repeat this process until there is no memory error for $M_{DP}$ (see line 15-22 in Algorithm~\ref{algo:SS_explosion}).
    \item Repeat the same procedure for the leakage power model of the CuV (see line 23-34 in Algorithm~\ref{algo:SS_explosion}) and the propagation delay model of the CuV ((see line 35-46 in Algorithm~\ref{algo:SS_explosion})). 
    \item After ensuring that there are no memory errors for each segment of $M_{DP}$, $M_{LP}$, $M_{D}$, and $M_{F}$, apply Algorithm~\ref{algo:sec_analysis} on each segment of $M_{DP}$, $M_{LP}$, $M_{D}$, and $M_{F}$.
\end{enumerate}

The segmentation of the model is closely linked with the decomposition of the LTL properties. In ForASec, the LTL properties are also generated by the verilog2smv tool. Therefore, when it translates the sub-models, it performs the decomposition of the properties. However, to incorporate the side-channel parameters, we modified these functional properties into side-channel parameter-based properties. It is important to note that the segmentation process is not automated, and the segmentation is done manually in the Verilog. The correctness of sub-models at the Verilog level is ensured by functional analysis. However, the correctness of sub-models at the SMV level is ensured by analyzing the switching activities of each node in CuV.

\textbf{Efficient HT-based security vulnerability analysis using side-channel parameters:} After segmentation, to further reduce the complexity of the verification, we propose to perform a proof by contradiction for each side-channel parameter-based and functional segments using respective LTL properties. In the proposed methodology, first, each segment of the CuV is functionally verified and verification is performed for each segment of the $M_{DP}$ (see line 6-10 in Algorithm~\ref{algo:sec_analysis}), the $M_{LP}$ (see line 11-14 in Algorithm~\ref{algo:sec_analysis}) and the $M_{D}$ (see line 15-19 in Algorithm~\ref{algo:sec_analysis}). Once the verification of all models is complete, the generated counterexamples are analyzed to identify the vulnerable nodes/gates/blocks in the system. These vulnerabilities can also be used to improve the CuV.

\begin{table*}[!t]
	\centering
	\caption{\textit{Structural Information of the ISCAS89 Benchmarks and the implemented HTs from trust-hub~\cite{salmani2013design,shakya2017benchmarking}.}}
	\label{benchmark_info}
	\scriptsize
	\resizebox{1\linewidth}{!}{
	\begin{tabular}{|c|c|c|c|c|c|c|c|c|c|c|c|c|c|c|c|c|c|c|c|}
        \hline
        \multicolumn{2}{|c|}{\multirow{3}{*}{\textbf{\begin{tabular}[c]{@{}c@{}}ISCAS \\ Benchmarks\end{tabular}}}} & \multirow{3}{*}{\textbf{Inputs}} & \multirow{3}{*}{\textbf{Outputs}} & \multicolumn{3}{c|}{\textbf{Gates}} & \multirow{3}{*}{\textbf{FF}} & \multicolumn{12}{c|}{\textbf{Size of HTs (Gates)}} \\ \cline{5-7} \cline{9-20} 
        \multicolumn{2}{|c|}{} &  &  & \multirow{2}{*}{\textbf{NANDs}} & \multirow{2}{*}{\textbf{NORs}} & \multirow{2}{*}{\textbf{NOTs}} &  & \multicolumn{4}{c|}{\textbf{T100}} & \multicolumn{3}{c|}{\textbf{T200}} & \multicolumn{5}{c|}{\textbf{T300}} \\ \cline{9-20} 
        \multicolumn{2}{|c|}{} &  &  &  &  &  &  & \textbf{NANDs} & \textbf{NORs} & \textbf{NOTs} & \textbf{FF} & \textbf{NANDs} & \textbf{NORs} & \textbf{NOTs} & \textbf{FF} & \textbf{NANDs} & \textbf{NORs} & \textbf{NOTs} & \textbf{FF} \\ \hline
        \textbf{B1} & \textbf{s27} & 4 & 1 & 2 & 6 & 5 & 3 & 1 & 1 & 1 &  & 1 & 1 & 1 &  & 1 & 1 & 1 &  \\ \hline
        \textbf{B2} & \textbf{s298} & 3 & 6 & 40 & 35 & 91 & 14 & 2 & 1 & 6 &  & 4 & 2 & 11 &  & 3 & 8 & 4 &  \\ \hline
        \textbf{B3} & \textbf{s344} & 9 & 11 & 62 & 39 & 112 & 15 & 2 & 1 & 6 &  & 4 & 2 & 11 &  & 2 & 9 & 2 & 3 \\ \hline
        \textbf{B4} & \textbf{s349} & 9 & 11 & 63 & 41 & 179 & 15 & 2 & 1 & 6 &  & 4 & 2 & 11 &  & 5 & 9 & 2 & 3 \\ \hline
        \textbf{B5} & \textbf{s382} & 3 & 6 & 41 & 58 & 94 & 21 & 2 & 1 & 6 & 1 & 4 & 2 & 11 &  & 3 & 8 & 4 &  \\ \hline
        \textbf{B6} & \textbf{s386} & 7 & 7 & 83 & 35 & 159 & 6 & 2 & 1 & 7 &  & 4 & 2 & 11 &  & 2 & 9 & 2 & 3 \\ \hline
        \textbf{B7} & \textbf{s400} & 3 & 6 & 47 & 59 & 94 & 21 & 4 & 5 & 8 & 1 & 8 & 5 & 15 & 1 & 2 & 9 & 2 & 3 \\ \hline
        \textbf{B8} & \textbf{s420} & 18 & 1 & 78 & 62 & 155 & 16 & 4 & 4 & 11 & 1 & 8 & 4 & 16 & 2 & 3 & 8 & 4 &  \\ \hline
        \textbf{B9} & \textbf{s444} & 3 & 6 & 71 & 48 & 89 & 21 & 4 & 9 & 6 & 1 & 8 & 5 & 15 & 1 & 2 & 9 & 2 & 3 \\ \hline
        \textbf{B10} & \textbf{s510} & 19 & 7 & 95 & 84 & 95 & 6 & 4 & 10 & 6 & 1 & 8 & 5 & 10 & 2 & 5 & 9 & 2 & 3 \\ \hline
        \textbf{B11} & \textbf{s526} & 3 & 6 & 88 & 63 & 136 & 21 & 4 & 4 & 11 & 1 & 4 & 5 & 8 & 3 & 3 & 8 & 4 &  \\ \hline
        \textbf{B12} & \textbf{s641} & 35 & 24 & 94 & 13 & 375 & 19 & 4 & 4 & 11 & 1 & 4 & 4 & 11 & 3 & 2 & 9 & 2 & 3 \\ \hline
        \textbf{B13} & \textbf{s713} & 35 & 23 & 122 & 17 & 365 & 19 & 2 & 1 & 6 & 1 & 4 & 9 & 6 &  & 5 & 9 & 2 & 3 \\ \hline
        \textbf{B14} & \textbf{s820} & 18 & 19 & 130 & 126 & 169 & 5 & 2 & 1 & 6 & 1 & 4 & 9 & 6 &  & 3 & 8 & 4 &  \\ \hline
        \textbf{B15} & \textbf{s832} & 18 & 19 & 132 & 130 & 167 & 19 & 2 & 1 & 7 &  & 4 & 4 & 11 & 3 & 2 & 9 & 2 & 3 \\ \hline
        \textbf{B16} & \textbf{s838} & 34 & 1 & 162 & 126 & 319 & 32 & 4 & 5 & 8 & 1 & 4 & 9 & 6 &  & 5 & 9 & 2 & 3 \\ \hline
        \textbf{B17} & \textbf{s953} & 16 & 23 & 163 & 148 & 169 & 29 & 4 & 4 & 11 & 1 & 4 & 9 & 6 &  & 3 & 8 & 4 &  \\ \hline
        \textbf{B18} & \textbf{s1196} & 14 & 14 & 237 & 151 & 370 & 18 & 2 & 1 & 6 & 1 & 4 & 2 & 11 &  & 2 & 9 & 2 & 3 \\ \hline
        \textbf{B19} & \textbf{s1238} & 14 & 14 & 259 & 169 & 217 & 18 & 2 & 1 & 7 &  & 4 & 2 & 11 &  & 5 & 9 & 2 & 3 \\ \hline
        \textbf{B20} & \textbf{s1423} & 17 & 5 & 161 & 229 & 501 & 74 & 4 & 5 & 8 & 1 & 4 & 2 & 11 &  & 3 & 8 & 4 &  \\ \hline
        \textbf{B21} & \textbf{s1488} & 8 & 17 & 350 & 200 & 653 & 6 & 4 & 4 & 11 & 1 & 8 & 5 & 15 & 1 & 2 & 9 & 2 & 3 \\ \hline
        \textbf{B22} & \textbf{s5378} & 35 & 49 & 0 & 1004 & 2014 & 179 & 4 & 9 & 6 & 1 & 8 & 4 & 16 & 2 & 5 & 9 & 2 & 3 \\ \hline
        \textbf{B23} & \textbf{s9234} & 19 & 22 & 1213 & 544 & 4956 & 228 & 4 & 10 & 6 & 1 & 8 & 5 & 15 & 1 & 3 & 8 & 4 &  \\ \hline
        \textbf{B24} & \textbf{s13207} & 13 & 121 & 1963 & 610 & 7004 & 669 & 7 & 4 & 11 & 1 & 8 & 5 & 10 & 2 & 2 & 9 & 2 & 3 \\ \hline
        \textbf{B25} & \textbf{s15850} & 77 & 150 & 2587 & 861 & 7443 & 534 & 7 & 4 & 11 & 1 & 8 & 5 & 15 & 1 & 5 & 9 & 2 & 3 \\ \hline
        \textbf{B26} & \textbf{s35932} & 35 & 320 & 11052 & 1152 & 9045 & 1728 & 51 & 2 & 64 & 1 & 3 & 12 & 4 &  & 3 & 8 & 4 &  \\ \hline
        \textbf{B27} & \textbf{s38417} & 28 & 106 & 6204 & 2505 & 17850 & 1636 & 3 & 9 & 4 &  & 3 & 12 & 11 &  & 2 & 9 & 2 & 3 \\ \hline
        \textbf{B28} & \textbf{s38584} & 12 & 278 & 7642 & 3806 & 15933 & 1452 & 6 & 2 & 8 &  & 21 & 1 & 22 &  & 5 & 9 & 2 & 3 \\ \hline
\end{tabular}}\vspace{-15pt}
\end{table*}
\section{Experimental Setup}
\begin{figure}[ht]
	\centering
	\includegraphics[width=0.85\linewidth]{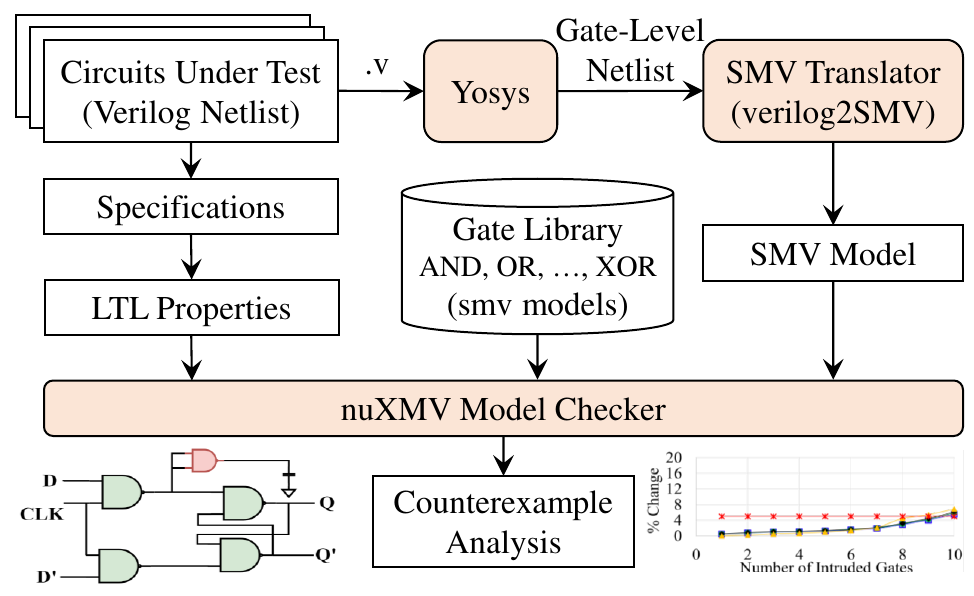}
	\caption{\textit{Experimental setup for evaluating the ForASec on multiple ISCAS benchmarks for trust-Hub HTs.}}
	\label{fig:exp_setup}
\end{figure}

To illustrate the practicability and usefulness of our ForASec framework, we evaluate it on a set of ISCAS89 benchmarks and some basic sequential circuits, like D-Flipflop (DFF), Shift register, and counter, with intruded gates (Table \ref{benchmark_info}) and trust-hub HT benchmarks. The following tools are used for experimentation, as shown in Fig. \ref{fig:exp_setup}: 
\begin{enumerate}[leftmargin=*]
	\item \textit{Yosys} is a tool that is used for the gate level translation of a given netlist or Verilog code \cite{wolf2016yosys}. 
	\item \textit{verilog2smv} tool is used to translate the gate level netlist into the corresponding smv mode \cite{irfan2016verilog2smv}.
	\item \textit{nuXmv Version 2.0} tool is used for model checking and performance analysis.
\end{enumerate} 

All the simulations are executed in Centos 7 on a computing machine with a Core i7 processor @3.4 GHz and 16 GB memory. For each benchmark, we consider two cases: 
\begin{enumerate}[leftmargin=*]
	\item \textbf{Case 1:} The malicious gates or circuits are inserted \textit{randomly} along within the IC paths at the input, output, and within critical and non-critical paths. These intrusions not only affect the capacitance of the nodes but also have a significant impact on the propagation delay.
	\item \textbf{Case 2:} The malicious gates are inserted \textit{parallel} to the IC paths. These intrusions only affect the capacitance of the nodes at which these circuits are inserted.
\end{enumerate} 

 
\subsection{Modeling and Analysis of Benchmark Circuits} 
The ForASec analysis of the implemented circuits is performed in two steps: (1) accuracy analysis of model translation and (2) vulnerability analysis against multiple intrusions. The correctness of the translated model is ensured by verifying all the functional, behavioral, and performance characteristics. 

\begin{figure}[!t]
	\centering
	\includegraphics[width=0.9\linewidth]{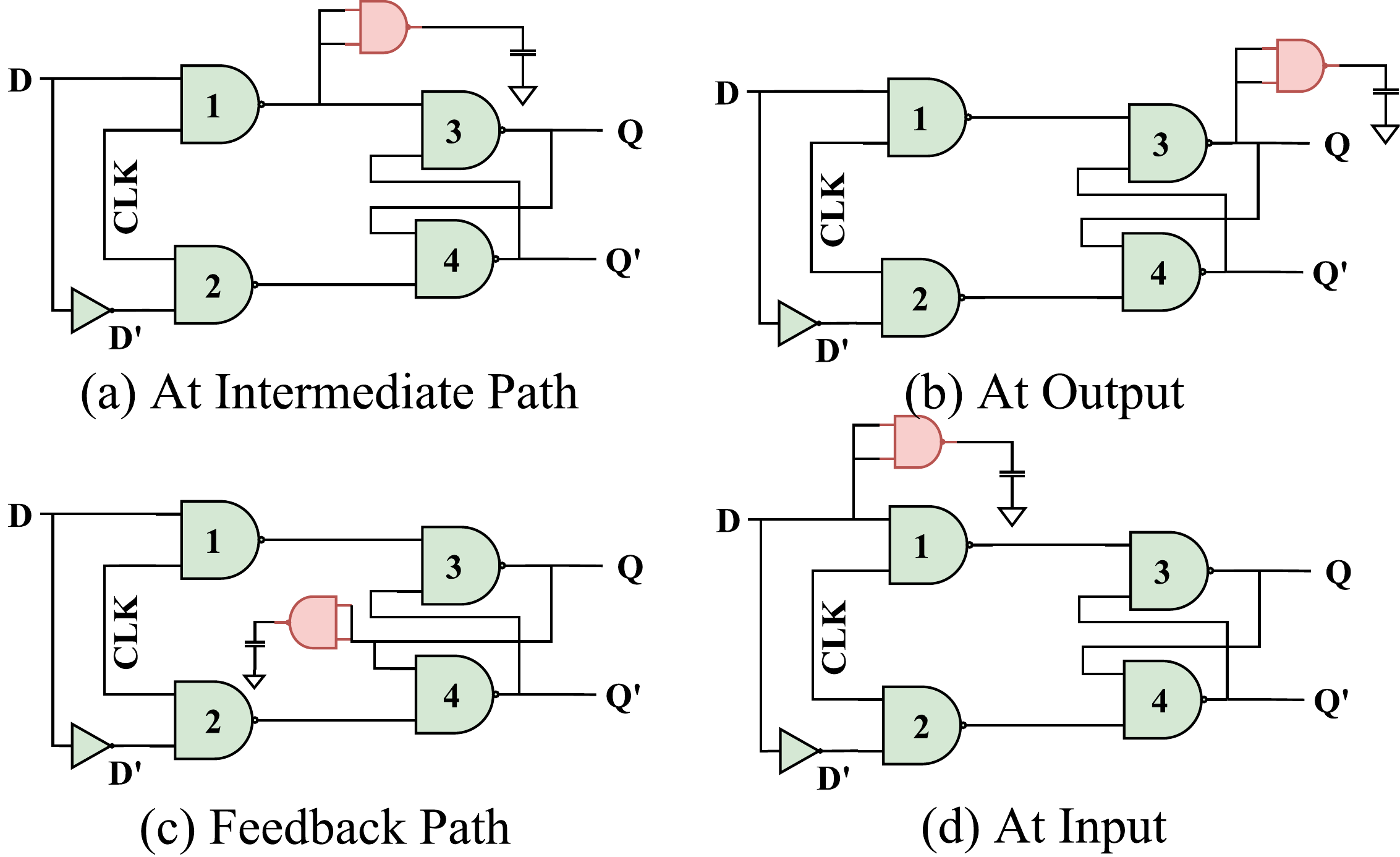}
	\caption{\textit{Some Examples of Implemented Intrusions in parallel to computational path at different locations of DFF.}}
	\label{fig:DFF}
\end{figure}

For example, in Fig. \ref{fig:DFF}, CLK-to-Q delay of a DFF is equivalent to the combined delay of NAND1 and NAND3, hold time of a DFF is equivalent to the inverters, delay and the setup time of DFF is equivalent to the combined delay of NAND3 and NAND4. Similarly, the dynamic power, functionality, and other characteristics are well defined in the literature. Therefore, to ensure the accuracy of the model translation, the translated model of DFF must fulfill all the properties mentioned above, design constraints, and characteristics. Similarly, we translated multiple ISCAS89 benchmarks along with basic sequential circuits, i.e., a DFF, shift registers, and counters, into their respective SMV models and verified them based on the properties mentioned above, design constraints, and characteristics.
\begin{figure*}
	\centering	\includegraphics[width=1\linewidth]{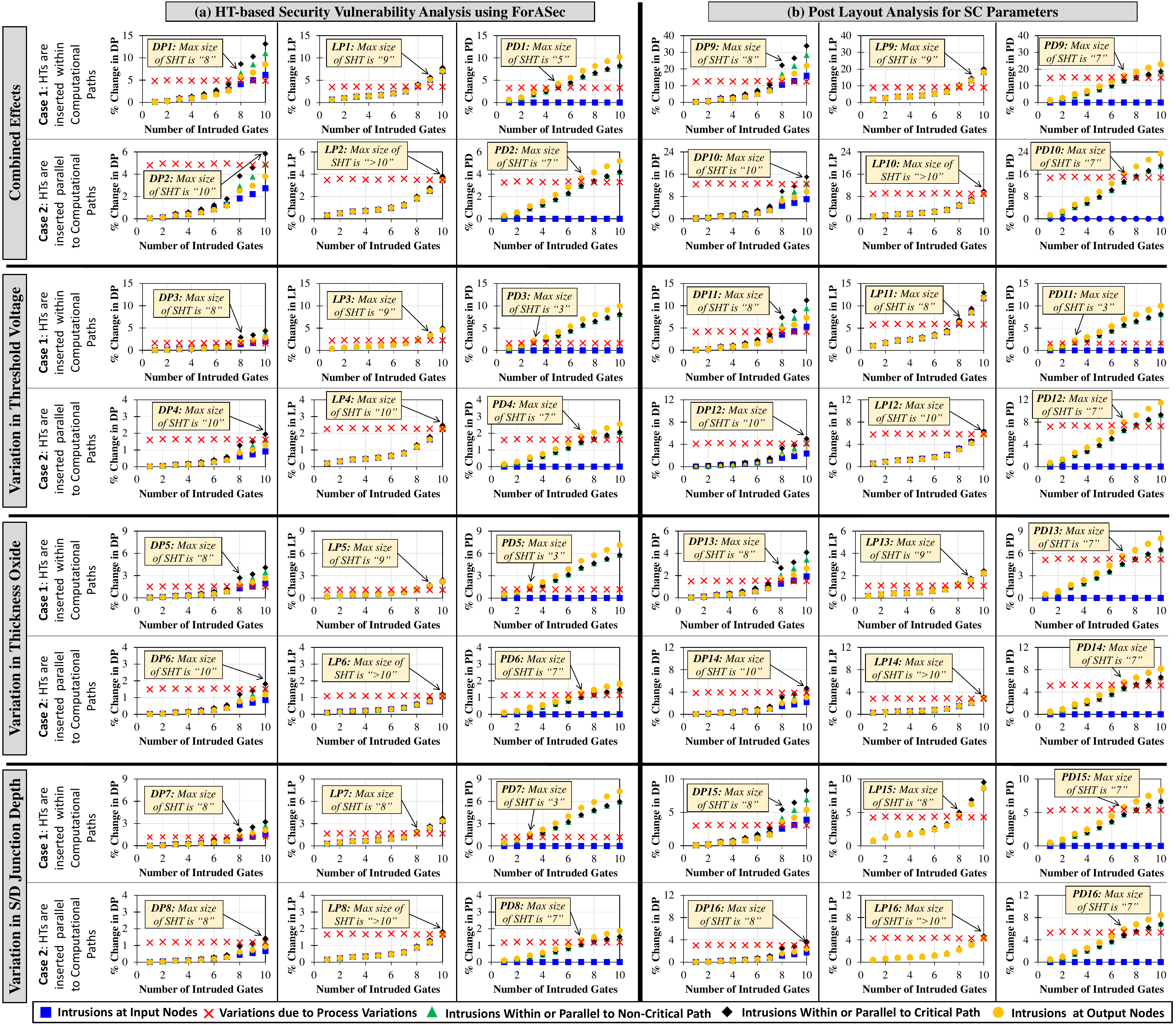}
	\caption{\textit{Analyzing the effects of process variations and intrusions by varying different parameters with Gaussian distribution (threshold Voltage, thickness oxide, junction depth, and their combined effect) and adding intruded gates at different positions, i.e., input (I/P), Non-Critical Path (NCP), Critical Path (CP) and output (O/P), on side-channel parameters in S349, ISCAS89 benchmark. Based on this analysis, we identify the maximum size (number of gates) of the SHT that remains stealthy with respect to change in dynamic power, change in leakage power, and change in propagation delay. All the analyses in (a) are obtained using ForASec, and all analyses in (b) are obtained  after synthesizing it using TSMC 22nm in Genus.}}\vspace{-15pt}
	\label{fig:in-line}
\end{figure*}
\section{HT-based Security Vulnerability Analysis} 
Once the accuracy is ensured, then the vulnerability of the translated model is analyzed by applying the effects of multiple intrusions. For instance, Fig. \ref{fig:DFF} shows multiple parallel intrusions in a DFF, which can affect the side-channel parameters due to the loading effect.

We evaluated basic sequential circuits and ISCAS89 benchmarks against single/multiple gate intrusions and trust-hub HT benchmarks, i.e., s35932-T100, s35932-T200, s35932-T300, s38417-T100, s38417-T200, s38417-T300, s38584-T100, s38584-T200, and s38584-T300. \textit{Due to space limitation, it is not possible to discuss the vulnerable point in all circuits, therefore, in this section, the analysis of \textbf{DFF} and ISCAS89 benchmarks, i.e., \textbf{s349, s35932, s38417, and s35854}, against single/multiple gate intrusions and trust-hub HT benchmarks, are presented as these are some the most complex circuits among the implemented ones.} Table \ref{benchmark_info} shows the structural information of the implemented benchmarks.

\subsection{Identifying vulnerable Nodes and Gates}
To identify the vulnerable nodes/gates in D flip-flop (DFF), we inserted HT gates at multiple locations in DFF, i.e., feedback paths/nodes, input nodes, output nodes, and intermediate paths/nodes, as shown in Fig. \ref{fig:DFF}. For example, if we insert the gate at the feedback node in DFF, the counterexample generated while verifying the Property~\ref{property:2} gives us the power consumption of each gate. Therefore, by comparing it with the un-intruded DFF, we can pinpoint the vulnerable gates. Similarly, counterexamples while verifying Properties~\ref{property:4}, we can get the value of abnormal value $C_{total}$ at the output of a gate, which contributes an additional delay. Hence, the nodes associated with that gate can be considered as potentially vulnerable nodes. Based on these counterexample analyses, we observed that the feedback node in DFF is more vulnerable than other positions because of higher switching activity, which makes this node slightly unstable and less vulnerable w.r.t. intrusion. Moreover, the change in the loading effect of the feedback node can cause the setup time violation, which may result in data corruption.

\begin{figure*}
	\centering	\includegraphics[width=1\linewidth]{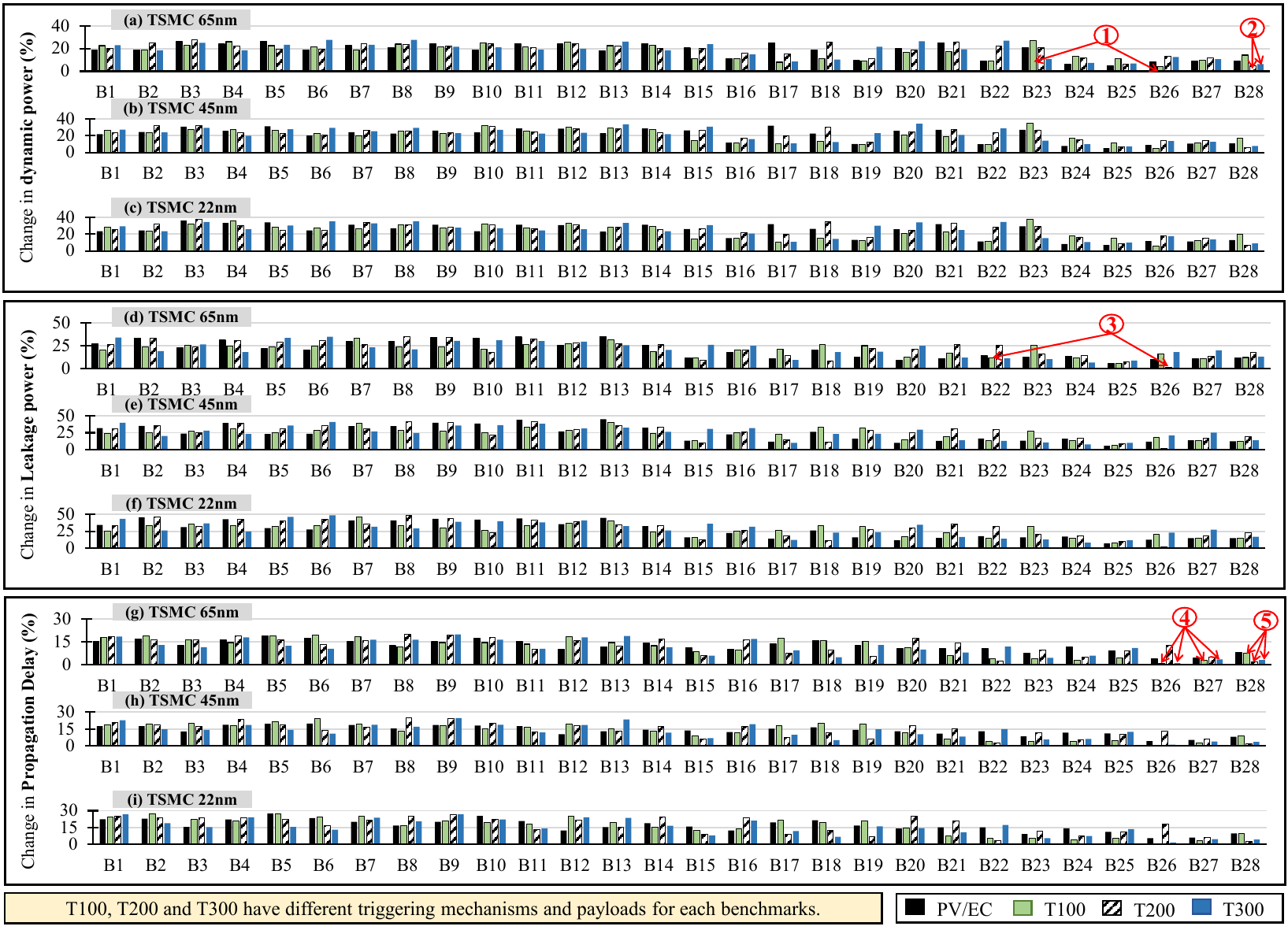}
	\caption{\textit{Effects of process variations/Environmental Changes (PV/EC) on side-channel parameters, i.e., propagation delay, dynamic and Leakage power, on different ISCAS89 benchmarks in the presence of trust-hub benchmarks. Note, in these experiments, process variations are modeled by non-deterministically choosing the thickness oxide ($t_{ox}$), S/D junction depth, and threshold voltage from a given range. We use 65nm, 45nm, and 22nm TSMC CMOS models to extract the ranges of these parameters. It also shows that for designing a stealthy HT, it must be inserted deeper (to minimize the triggering probabilities) in the circuits and inserted in parallel (to minimize the effect on delay and dynamic power) to the computational path. In this analysis, we observed that different technologies have an impact on side-channel parametric behavior, but it does not affect the outcome of the security vulnerability analysis.}}
	\label{fig:EA_trust-hub}\vspace{-15pt}
\end{figure*}

\subsection{Identifying the maximum size of SHT} 
For this vulnerability analysis, we intruded the S349 at multiple locations with different sizes (the number of intruded gates). Fig. \ref{fig:in-line} shows the effects of intrusion on the side-channel parameters but \textbf{the effects of process variations} can overshadow these variations. Therefore, before intruding the circuits, we analyzed the effects of process variations on threshold voltage, thickness oxide, and junction depth, shown in Fig. \ref{fig:in-line} and analyzed their individual and combined effects, which exhibit the following key observations:
\begin{enumerate}[leftmargin=*]
	
	\item \textbf{The effect on dynamic power} due to the process variations overshadows the effects of intrusions if the maximum size of SHT is 8 (i.e., 2\% of S349), as shown in Fig. \ref{fig:in-line}(a) (see label DP1). However, beyond that, the effects of intrusions are dominant and vary at different locations. For example, the dynamic power is affected by almost 11\% when 10 gates are intruded within the critical path. However, it is affected by almost 5\%, when 10 gates are intruded at the input locations. Similarly, effects can be observed when analyzing the impact of process variations on key parameters, i.e., threshold voltage, thickness oxide, and junction depth. The reason behind this behavior is that, in S349, \textbf{there are no DFFs at the input}, but the intrusion at intermediate stages involves effects in the feedback path of DFF, which starts the unwanted switching activity. Similar behavior can be observed in the case of parallel gate intrusion (see label DP2).  
	
	\item \textbf{The effect on leakage power} due to the process variations overshadows the effects of intrusions if the HT size is less than or equal to 9 gates (2.5\% of S349), as shown in Fig. \ref{fig:in-line}(b) (see labels LP1). The reason behind this behavior is that most of the physical parameters are affected by process variations. Similarly, while observing the impact of process variations on key parameters, i.e., threshold voltage, thickness oxide, and junction depth, individually, the maximum size of SHT is observed between 9 and 11 (see labels LP2, LP3, LP4, LP5, LP6, LP7, and LP8).
	
	\item \textbf{The effect on propagation delay} due to gate intrusions is very dominant, especially, when \textit{the intrusions are within the computational path} and maximum size of the SHT varies from 3 to 5, as shown in Fig. \ref{fig:in-line}(a) (see labels PD1, PD3, PD5, and PD7). However, if \textit{intrusions are in parallel} then the effect remains within the range of the process variations and the maximum size of the SHT is relatively large, and for all cases, its value is 7, as shown in Fig. \ref{fig:in-line}(a) (see label PD2, PD4, PD6, and PD8).     
\end{enumerate} 

\begin{figure*}[!t]
	\centering	\includegraphics[width=1\linewidth]{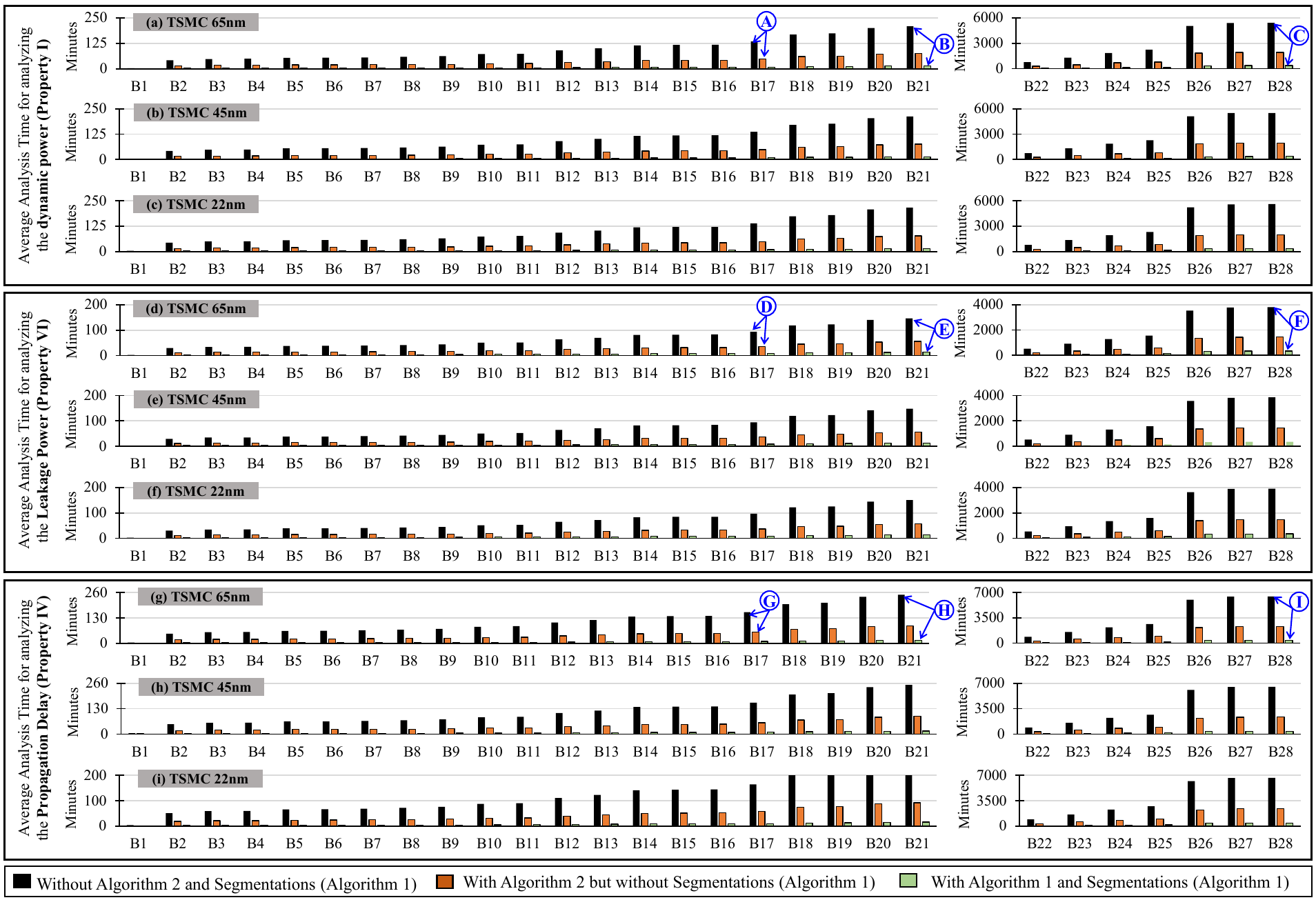}
	\caption{\textit{Average analysis time [minutes] for analyzing the different ISCAS89 benchmarks with and without proposed analysis algorithm~\ref{algo:sec_analysis} and segmentation-based techniques. The analysis shows that algorithm~\ref{algo:sec_analysis} with segmentation (Algorithm~\ref{algo:SS_explosion}) can provide approximately 11x to 16x speed in analysis time compared to state-of-the-art model checking-based techniques~\cite{imran2016,abbassi2019using,abbassi2018mcsevic}. In this analysis, we also observed that the different technologies do not significantly impact the analysis time. Note, the analysis time is measured using the ``time'' command of nuXmV. All the verification experiments are performed in Centos 7 on a machine with a Core i7 processor @3.4 GHz and 16 GB memory. }}
	\label{fig:Analysis_time_mins}\vspace{-10pt}
\end{figure*}
As a consequence, it can be concluded that \textit{input nodes are the most vulnerable point in the sequential circuits} because the effects of such intrusions can be overshadowed by process variations, except for DFF in which the most vulnerable points are feedback nodes.

To illustrate the effectiveness of ForASec-based analysis, we also perform the post-layout analysis in Cadence Genus using 22nm TSMC technology. The comparison of these analyses shows that in all cases, the side-channel behavior of S349 is similar, as shown in Fig. \ref{fig:in-line}(b). We also observed that in most cases, the maximum size of SHT is the same. However, in a few cases, the maximum size is slightly higher than the estimated in ForASec. For example, case 1 of propagation delay (compare label PD1 with PD9, and PD5 with PD13) in combined effects experiments and case 1 of leakage power (compare label LP3 with LP11) in the experiments when the $V_{th}$ is varied.

\textbf{ISCAS89 benchmark against HT benchmarks:} 
To illustrate the effectiveness of the proposed methodology, we evaluated ForAsec for all ISCAS89-benchmarks using different ASIC technologies, i.e., 65nm, 45nm, and 22nm, for technology parameters and process variation models, see Figures~\ref{fig:EA_trust-hub} and \ref{fig:Analysis_time_mins}.
  
In most cases, the change in side-channel dynamic power, leakage power, and propagation delay are distinguishable from PV/EC. However, in the case of some HT benchmarks, the change in dynamic power (see labels 1 and 2 in Figure~\ref{fig:EA_trust-hub}), leakage power (see label 3 in Figure~\ref{fig:EA_trust-hub}) and propagation delay (see labels 4 and 5 in Figure~\ref{fig:EA_trust-hub}) is with the PV/EC cap. Based on these observations, we can observe that some HTs can remain stealthy with respect to side-channel parameters in ISCAS89 benchmarks. Moreover, this analysis shows that different technologies have an impact on side-channel parametric behavior, but it does not affect the outcome of the security vulnerability analysis.

The analyses presented in Figs.~\ref{fig:in-line} and \ref{fig:EA_trust-hub} also show that for designing a stealthy HT, it must be inserted deeper (to minimize the triggering probabilities) in the circuits and inserted in parallel (to minimize the effect on delay and dynamic power) to the computational path.

\subsection{Performance Analysis}\label{performance}
\begin{table}[!t]
	\centering
	\caption{\textit{Performance Comparison with respect to Analysis Time (T(s)) and required Memory(M(GB)) for 8 gates intrusion (with computational path) in S349, ISCAS89 benchmark.}}
	\label{tab:time}
	\footnotesize
	\resizebox{1\linewidth}{!}{
		\begin{tabular}{|c|c|c|c|c|c|c|}
			\hline
			\multirow{2}{*}{\textbf{Parameter}} & \multicolumn{1}{c|}{\multirow{2}{*}{\textbf{Location}}} & \multicolumn{1}{c|}{\multirow{2}{*}{\textbf{\# of CEs}}} & \multicolumn{2}{c|}{\textbf{\begin{tabular}[c]{@{}c@{}}Without Algorithm \ref{algo:sec_analysis}\end{tabular}}} & \multicolumn{2}{c|}{\textbf{\begin{tabular}[c]{@{}c@{}}With Algorithm \ref{algo:sec_analysis}\end{tabular}}} \\ \cline{4-7} 
			& \multicolumn{1}{c|}{} & \multicolumn{1}{c|}{} &\textbf{T(s)} & \textbf{M(GB)} &  \textbf{T(s)} & \textbf{M(GB)} \\ \hline
			\multirow{4}{*}{\textbf{\begin{tabular}[c]{@{}c@{}}Switching \\ Power\\ Property I\end{tabular}}} & \textbf{Input} & 221 & 39579 & 16 & 13853 & 16   \\ \cline{2-7} 
			& \textbf{Output} & 304 & 47978 & 16  & 16792 & 16  \\ \cline{2-7} 
			& \textbf{CP} & 121 & 21642 & 16 & 7575 &  13.94  \\ \cline{2-7} 
			& \textbf{NCP} & 102 & 17452 & 16 & 6108 &  12.7  \\ \hline
			\multirow{4}{*}{\textbf{\begin{tabular}[c]{@{}c@{}}Leakage\\ Power\\ Property VI\end{tabular}}} & \textbf{Input} & 29 & 4521 & 16 & 1718 & 8.02   \\ \cline{2-7}
			& \textbf{Output} & 31 & 4684 & 16 & 1780 &  8.01  \\ \cline{2-7} 
			& \textbf{CP} & 179 & 32574 & 16 & 12447 &  15.17  \\ \cline{2-7}
			& \textbf{NCP} & 164 & 30167 & 16 & 11463 &  14.9  \\ \hline
			\multirow{4}{*}{\textbf{\begin{tabular}[c]{@{}c@{}}Propagation\\ Delay\\ Property IV\end{tabular}}} & \textbf{Input} & 269 & 44579 & 16 & 16450  & 16   \\ \cline{2-7} 
			& \textbf{Output} & 356 & 55897 & 16 & 20626 & 16   \\ \cline{2-7} 
			& \textbf{CP} & 124 & 23591 & 16 & 8705 & 13.8   \\ \cline{2-7} 
			& \textbf{NCP} & 103 & 18694 & 16 & 6998 & 12.74   \\ \hline
	\end{tabular}}
\end{table}
\begin{figure*}[!t]
    \centering	\includegraphics[width=1\linewidth]{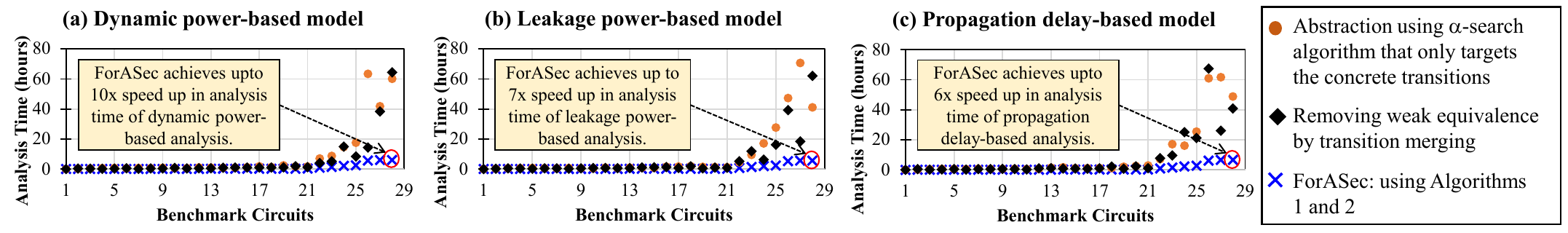}
    \caption{\textit{Average analysis time [hours] for analyzing the different ISCAS89 benchmarks using the proposed methodology and traditional techniques to avoid state-space explosion, i.e., $\alpha$-search and remove weak equivalences. The analysis shows that the proposed methodology not only addresses the state-space explosion but also acquires up to 10x speedup in the analysis time. Note, the analysis time is measured using the ``time'' command of nuXmV model checker. All the verification experiments are performed in Centos 7 on a machine with a Core i7 processor @3.4 GHz and  16 GB memory.}}\vspace{-15pt}
    \label{fig:comp_SoA}
\end{figure*}
For a single violation in a property, multiple counterexamples exist depending upon intruded gates, their size, and locations. However, it increases the time required by the nuXmv model checker to verify a single property, and it also increases the size of the state-space model. Table~\ref{tab:time} shows the number of counterexamples (CE), their analysis time (T) in seconds (s), and memory (M) acquired in gigabytes (GB) for the 8 NAND gates intrusion (within a computational path) in the ISCAS89 benchmark S349. It shows that a small benchmark with 349 gates requires a huge amount of memory and time. Therefore, ForASec generates and analyzes all possible counterexamples using Algorithm~\ref{algo:sec_analysis} by verifying one SC parameter at a time. Table~\ref{tab:time} shows that with our proposed algorithm, the analysis time is reduced by 66\%. For example,  in the case of input intrusions, the time required to verify Property I without Algorithm~\ref{algo:sec_analysis} is 39579s, but the time required to verify Property I with Algorithm~\ref{algo:sec_analysis} is 13853s. 

\subsection{Average Analysis Time:} \label{sec:MAT}
To evaluate the effectiveness of the segmentation-based approach with and without Algorithm 1, we analyze all the ISACAS89 benchmarks. Fig.~\ref{fig:Analysis_time_mins} shows the average analysis time to verify the different properties, i.e., dynamic power (Property I), leakage power (Property VI), and propagation delay (Property IV), of several ISCAS89 benchmarks with HT benchmarks from trust-hub. By analyzing these results, we made the following key observations:
    
    \begin{enumerate}[leftmargin=*]
        \item For all the implemented ISCAS89 benchmarks, the security analysis using only Algorithm~\ref{algo:sec_analysis}, i.e., verification using the inversion of LTL properties, reduces the analysis time. For example, the analysis time for B17 reduces by 2.8x (see label A in Figure~\ref{fig:Analysis_time_mins}) in dynamic power-based analysis, by 2.6x (see label D in Figure~\ref{fig:Analysis_time_mins}) in leakage power-based analysis, and 4.3x (see label G in Figure~\ref{fig:Analysis_time_mins}) in propagation delay based analysis. 
        \item Similarly, for all the implemented ISCAS89 benchmarks, the security analysis using segmentation (Algorithm~\ref{algo:SS_explosion}) along with Algorithm~\ref{algo:sec_analysis} further reduces the analysis time. For example, the analysis time for B21 reduces by 15x (see label B in Figure~\ref{fig:Analysis_time_mins}) in dynamic power-based analysis, by 11x (see label E in Figure~\ref{fig:Analysis_time_mins}) in leakage power-based analysis and 17x (see label H in Figure~\ref{fig:Analysis_time_mins}) in propagation delay based analysis. It is important to note that a similar speedup is observed in all the ISCAS89 benchmarks. For example, the analysis time for B28 reduces by 15.2x (see label B in Figure~\ref{fig:Analysis_time_mins}) in dynamic power-based analysis, by 11.2x (see label F in Figure~\ref{fig:Analysis_time_mins}) in leakage power-based analysis and 16.3x (see label I in Figure~\ref{fig:Analysis_time_mins}) in propagation delay based analysis.
        \item We also observed that different technologies have an impact on side-channel parametric behavior, but it does not affect the outcome of the security vulnerability analysis.
    \end{enumerate}
\textit{Therefore, based on these observations, we can conclude that unlike the traditional MC-based analysis, ForASec can verify the complex and large sequential circuits without affecting the correctness of the model and interdependencies between different gates/components, and it provides approximately 11x to 16x speed in analysis time compared to state-of-the-art model checking-based techniques~\cite{imran2016,abbassi2019using,abbassi2018mcsevic}. Moreover, the comparison with post-layout synthesis shows the robustness of ForASec against process variations.}

\section{Comparison with traditional techniques for state-space reduction}\label{sec:SoA}
To illustrate the effectiveness of the proposed technique, we applied ForASec and other traditional state-space reduction techniques, i.e., $\alpha$-search algorithm and removing weak equivalence, on the ISCAS89 benchmarks circuits (see Figure~\ref{fig:comp_SoA}). From this analysis, we made the following key observations:  
\begin{enumerate}[leftmargin=*]
    \item In all the analyses, ForASec has the lowest analysis time. The reason for this behavior is that the segmentation technique of ForASec considers the available RAM for computing the size of the segment. Moreover, each resultant segment has only one output to reduce the output feedback loops to one.
    \item ForAsec achieved up to 10x, 7x, and 6x speedup for analyzing the DP-based models, LP-based models, and PD-based models, respectively. 
\end{enumerate}

\section{Conclusion}\label{conclusion}
In this paper, we propose a novel formal framework (ForASec), based on the side-channel parameter analysis, to analyze/identify the vulnerabilities in sequential circuits against multiple types of intrusions at different locations while considering the effects of process variations on the technology parameters. Moreover, this framework addresses the state-space problem of the traditional MC-based approaches by dividing the state-space into several sub-state-spaces. For illustration, the proposed methodology has been applied to multiple ISCAS89 benchmarks and standard sequential circuits for different ASIC technologies, i.e., 65nm, 45nm, 22nm. The experimental results show that ForASec can analyze the complex and large sequential circuits and provides approximately 6x to 10x speedup in analysis time compared to state-of-the-art MC-based techniques. 
\section*{Acknowledgment}
This work is supported in parts by the Austrian Research Promotion Agency (FFG) and the Austrian Federal Ministry for Transport, Innovation, and Technology (BMVIT) under the “ICT of the Future” project, IoT4CPS: Trustworthy IoT for Cyber-Physical Systems.

\footnotesize
\bibliographystyle{IEEEtran}
\bibliography{bib/bibliography}
\appendix
\subsection{Computing the Number of Test Patterns}\label{Sec:TP}
\begin{figure}[h]
    \centering	\includegraphics[width=1\linewidth]{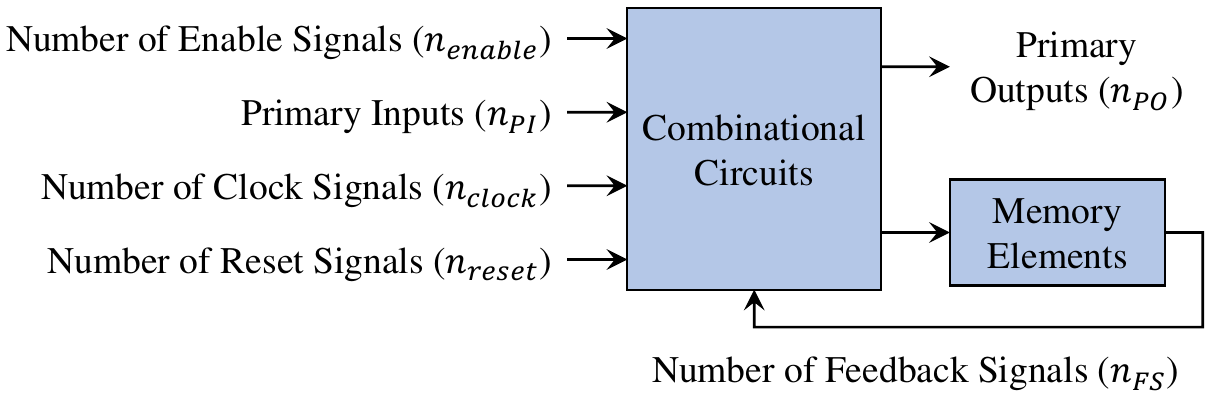}
    \caption{\textit{A generic block diagram of a sequential circuit that shows all the inputs, outputs and feedback signals.}}
    \label{fig:block_SC}
\end{figure}

The output of a combinational circuit depends on the current state of the input combinations. Therefore, the number of test patterns required for analysis with 100\% coverage is $2^{n_{inputs}}$. However, the output of the sequential circuit depends on the current and previous state of inputs, i.e., primary inputs, clock signals, reset signals, enable signals, and feedback signals~\cite{mano2013digital}. For example, Fig.~\ref{fig:block_SC} shows a generic block diagram of a sequential circuit. For 100\% coverage in test pattern-based analysis, the number of test patterns (TP) also depends on the current and previous state of the inputs.
\begin{table}[!t]
	\centering
	\caption{\textit{List of Abbreviations}}
	\label{Abbreviations}
	\resizebox{1\linewidth}{!}{
	\begin{tabular}{ll}
		\hline
		Abbreviation & Representation \\ \hline
		\textit{HT} & \textit{Hardware Trojans} \\
		\textit{SHT, AHT} & \textit{Stealthy and Active Hardware Trojans} \\
		\textit{SE} & \textit{Soft Errors} \\
		\textit{Seq, Comb} & \textit{Sequential and Combinational Circuits} \\
		\textit{PV} & \textit{Process Variations} \\
		\textit{SC} & \textit{Side-Channel Parameters} \\
		\textit{$R_{on}$} & \textit{Switch on Resistance} \\
		\textit{$t_{ox}$} & \textit{Oxide Capacitance} \\
		\textit{$\mu$} & \textit{Carrier (Electron ($\mu_{n}$) / Holes $\mu_{p}$) Mobilities} \\
		\textit{$C_{g}$, $C_{d}$, $C_{s}$} & \textit{Gate, Drain and Source Capacitances} \\
		\textit{$C_{diff}$} & \textit{Internal Diffusion Capacitance} \\
		\textit{$C_{load}$} & \textit{Load Capacitance} \\
		\textit{$C_{dmin}$} & \textit{Minimum Drain Diffusion Capacitance} \\
		\textit{$C_{jbd}$} & \textit{\begin{tabular}[l]{@{}l@{}}Capacitance per unit area between body and bottom \\ of the drain\end{tabular}} \\
		\textit{$C_{jbsdw}$} & \textit{\begin{tabular}[l]{@{}l@{}}Capacitance per unit length of the junction between \\ body and side walls of the drain.Works\end{tabular}} \\
		\textit{$C_{smin}$} & \textit{Minimum Source Diffusion Capacitance} \\
		\textit{$P_{switching}, DP$} & \textit{Switching/Dynamic Power} \\
		\textit{$\alpha$} & \textit{Switching Activity} \\
		\textit{$C_{total}$} & \textit{Total Capacitance that is (dis)charged in a transition} \\
		\textit{$f$} & \textit{Operating Frequency} \\
		\textit{$Vdd, V_{th}$} & \textit{Supply Voltage and Threshold Voltage} \\
		\textit{$P_{leakage}, LP$} & \textit{Leakage Power} \\
		\textit{$I_{leakage}$} & \textit{Leakage Current} \\
		\textit{$W$} & \textit{Gate Width} \\
		\textit{$L$} & \textit{Effective Channel Length} \\
		\textit{$\phi{_t}$ } & \textit{Thermal voltage} \\
		\textit{$\mathit{\sigma}$} & \textit{Drain Induced Barrier Lowering (DIBL) Factor} \\
		\textit{$t_{delay}, D, PD$} & \textit{Propagation Delay} \\
		\textit{DFF} & \textit{D Flip-Flop} \\
		\textit{CE} & \textit{Counterexample} \\
		\textit{$C_{GDO}, C_{GSO}$} & \textit{Gate-Drain and Gate-Source Overlap} Capacitances \\ 
		\textit{SMV} & \textit{Symbolic Model Verification} \\
		\textit{CP} & \textit{Intrusions are in critical path} \\
		\textit{NCP} & \textit{Intrusions are in non-critical path} \\
		\textit{CuV} & \textit{Circuit under Verification} \\
		\textit{MC} & \textit{Model Checking} \\
		\hline		
	\end{tabular}}
\end{table}
\begin{itemize}[leftmargin=*]
    \item \textit{$TP_{inputs}$}: Based on this assumption, in the motivational analysis, we compute the number of TP required to verify all input combinations using the following set of equations: 
    \begin{equation}
    \baselineskip=0pt
	\scriptsize
        TP_{inputs}= 2^{n_{inputs\_previous}}\times 2^{n_{inputs\_current}} 
        \label{eq:tp_inputs}
    \end{equation}
        
    Since the number of inputs does not change and $TP_{inputs}$ depends on the previous and current state of the same inputs. Therefore, we can replace $n_{inputs\_previous}$ and $n_{inputs\_current}$ with $n_{inputs}$ in Equation~\ref{eq:tp_inputs}. 
    \begin{equation}
    \baselineskip=0pt
	\scriptsize
        TP_{inputs}= 2^{n_{inputs}}\times 2^{n_{inputs}} = 2^{(2 \times n_{inputs})}
        \label{eq:tp_inputs_1}
    \end{equation}
    \begin{equation}
    \baselineskip=0pt
	\scriptsize
        n_{inputs} = n_{PI} + n_{FS} + n_{clock} + n_{reset} + n_{enable}  
        \label{eq:n_inputs}
    \end{equation}
    Where $n_{PI}$, $n_{FS}$ $n_{clock}$, $n_{reset}$ and $n_{enable}$ are the number of primary inputs, number of feedback signals, number of clock signals, number of reset signals and number of enable signals, respectively.  
    
    \item \textit{$TP_{nodes}$} Similar to \textit{$TP_inputs$}, the analysis with respect to nodes for sequential circuits also depends on the current and previous state of the nodes. Therefore, we compute the \textit{number of TP required to verify all nodes} using the following equation:
    \begin{equation}
    \baselineskip=0pt
	\scriptsize
        TP_{nodes} = 2^{n_{nodes}}\times 2^{n_{nodes}} = 2^{(2 \times n_{nodes})}
        \label{eq:tp_nodes}
    \end{equation}
    Where $n_{nodes}$ represents the total number of nodes in a sequential circuit. 
        
    \item $TP_{gates}$: In the case of analysis with respect to gates, the total test pattern depends on the number of gates, the number of inputs of a gate, and the current and previous state of the inputs of a gate. Therefore, we compute the number of TP required to verify all gates using following set of equations:
    \begin{equation}
    \baselineskip=0pt
	\scriptsize
        TP_{gates} = \Big \{\sum\limits_{i=1}^{n_{g}} 2^{g(i)}\Big \} \times \Big \{\sum\limits_{i=1}^{n_{g}} 2^{g(i)}\Big \} 
            = \Big \{\sum\limits_{i=1}^{n_{g}} 2^{g(i)}\Big \}^2
        \label{eq:tp_gates}
    \end{equation}
    Where $n_{g}$ is the number of number of gates in a sequential circuit and $g(i)$ is the number of inputs in an \textit{``ith''} gate.
        
    \end{itemize}

\begin{IEEEbiography}[{\includegraphics[width=1in,height=1.25in,clip,keepaspectratio]{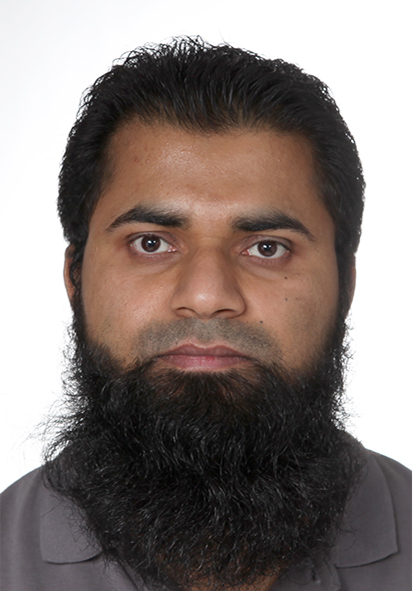}}]{Faiq Khalid}(GS'18-M'21) received his M.S. degree in electrical engineering and his B.E. degree in electronics engineering from the National University of Sciences and Technology (NUST), Pakistan, in 2016 and in 2011, respectively. He is currently pursuing his Ph.D. degree in hardware security and machine learning security at Technische Universit{\"a}t Wien (TU Wien), Vienna, Austria. He is a recipient of the Quaid-e-Azam Gold Medal for his academic achievements and the Richard Newton Young Fellowship Award at DAC 2018. His research interests include formal verification of embedded systems, hardware design security, and security for machine learning systems.   
\end{IEEEbiography}
\begin{IEEEbiography}[{\includegraphics[width=1in,height=1.25in,clip,keepaspectratio]{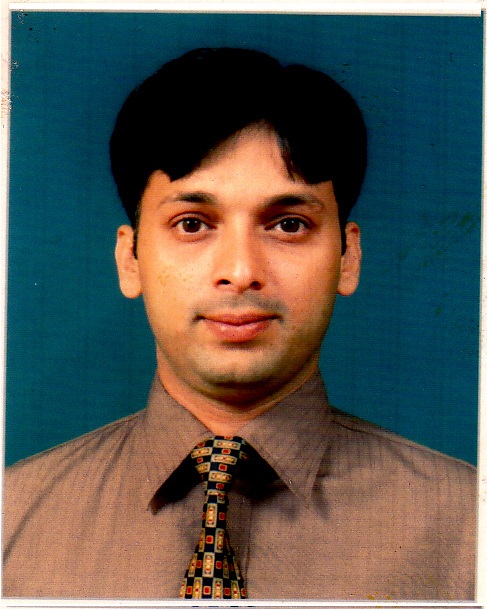}}]{Imran Hafeez Abbassi} received BSE degree in Avionics Engineering from College of Aeronautical Engineering (CAE), NUST in 2002, his MS degree in Information Security and PhD degree in Electrical Engineering from NUST, Pakistan in 2011 and 2019, respectively. Currently, he is an Associate Professor in CAE since April 2019. His research interests include hardware security, cryptography, formal verification, machine learning and embedded system design. His industrial experience spans over more than 10 years, during which he designed and developed diverse range of avionics systems for aerial platforms. He is also an awardee of PhD scholarship from Higher Education Commission, Pakistan.
\end{IEEEbiography}
\begin{IEEEbiography}[{\includegraphics[width=1in,height=1.25in,clip,keepaspectratio]{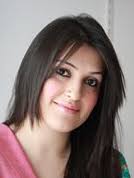}}]{Semeen Rehman} is currently with the Technische Universität Wien (TU Wien), Faculty of Electrical Engineering, as a tenure-track Assistant Professor. In October 2020, she received her habilitation in the area of Embedded Systems from TU Wien. Before that, she was a Postdoctoral Researcher with the Technische Universität Dresden (TU Dresden) and Karlsruhe Institute of Technology (KIT), Germany, since 2015. In July 2015, she received her Ph.D. from KIT, Germany. Her main research interests include dependable systems, cross-layer design for error resiliency with a focus on run-time adaptations, emerging computi  ng paradigms, such as approximate computing, hardware security, energy-efficient computing, embedded systems, MPSoCs, Internet of Things, and Cyber-Physical Systems. She received the CODES+ISSS 2011 and 2015 Best Paper Awards, DATE 2017 Best Paper Award Nomination, several HiPEAC Paper Awards, Richard Newton Young Student Fellow Award at DAC 2015, and Research Student Award at KIT in 2012. She has served as the TPC track chair for the ISVLSI 2020 conference and on the TPC of multiple premier conferences on design automation and embedded systems (DAC, DATE CASES, ASP-DAC). She has (co-)chaired sessions at the DATE 2019, 2018, and 2017 conferences.
\end{IEEEbiography}
\begin{IEEEbiography}[{\includegraphics[width=1in,height=1.2in,clip]{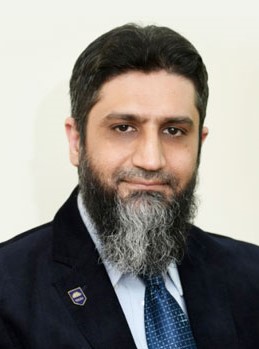}}]{Osman Hasan} received the M.Eng. and Ph.D. degrees from Concordia University, Montreal, Canada, in 2001 and 2008, respectively. Currently, he is an Associate Professor at the National University of Science and Technology (NUST), Islamabad, Pakistan. He is the founder and director of System Analysis and Verification (SAVe) Lab at NUST, which mainly focuses on the design and formal verification of safety-critical systems, including e-health and digital systems. He has received several awards and distinctions, including the Pakistan’s Higher Education Commission’s Best University Teacher (2010) and Best Young Researcher Award (2011) and the President’s gold medal for the best teacher of the University from NUST in 2015. Dr. Hasan is a senior member of IEEE, member of the ACM, Association for Automated Reasoning (AAR) and the Pakistan Engineering Council.  
\end{IEEEbiography}
\begin{IEEEbiography}[{\includegraphics[width=1in,height=1.2in,clip]{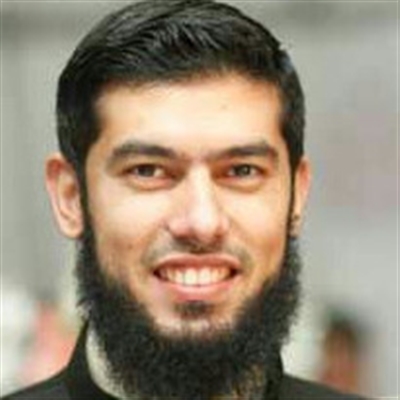}}]{Awais Mehmood Kamboh} (S’06–M’11-SM’17) received his B.E. degree in electrical engineering from National University of Sciences and Technology, Islamabad, Pakistan, in 2003, his M.S. degree in electrical engineering: systems from the University of Michigan, Ann Arbor, Michigan, USA in 2006, and his Ph.D. degree in electrical engineering from Michigan State University, East Lansing, Michigan, USA, in 2010. He joined the School of Electrical Engineering and Computer Science (SEECS), National University of Sciences and Technology (NUST), Islamabad, Pakistan. In 2011. Dr. Kamboh joined College of Computer Science and Engineering (CCSE), University of Jeddah (UJ), Jeddah, Kingdom of Saudi Arabia in 2019 as an Associate Professor. His research interests include mixed-signal integrated-circuits, biomedical signal processing, and brain computer interfaces. 
\end{IEEEbiography}
\begin{IEEEbiography}[{\includegraphics[width=1in,height=1.25in,clip,keepaspectratio]{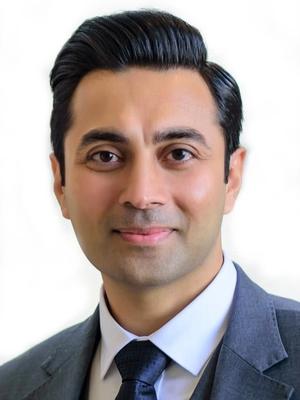}}]{Muhammad Shafique}(M'11 - SM'16) received the Ph.D. degree in computer science from the Karlsruhe Institute of Technology (KIT), Germany, in 2011. Afterwards, he established and led a highly recognized research group at KIT for several years as well as conducted impactful R\&D activities in Pakistan. In Oct.2016, he joined the Institute of Computer Engineering at the Faculty of Informatics, Technische Universität Wien (TU Wien), Vienna, Austria as a Full Professor of Computer Architecture and Robust, Energy-Efficient Technologies. Since Sep.2020, he is with the Division of Engineering, New York University Abu Dhabi (NYU AD), United Arab Emirates.

His research interests are in brain-inspired computing, AI \& machine learning hardware and system-level design, energy-efficient systems, robust computing, hardware security, emerging technologies, FPGAs, MPSoCs, and embedded systems. His research has a special focus on cross-layer analysis, modeling, design, and optimization of computing and memory systems. The researched technologies and tools are deployed in application use cases from Internet-of-Things (IoT), smart Cyber-Physical Systems (CPS), and ICT for Development (ICT4D) domains.

Dr. Shafique has given several Keynotes, Invited Talks, and Tutorials, as well as organized many special sessions at premier venues. He has served as the PC Chair, Track Chair, and PC member for several prestigious IEEE/ACM conferences. 
Dr. Shafique holds one U.S. patent has (co-)authored 6 Books, 10+ Book Chapters, and over 200 papers in premier journals and conferences. He received the 2015 ACM/SIGDA Outstanding New Faculty Award, AI 2000 Chip Technology Most Influential Scholar Award in 2020, six gold medals, and several best paper awards and nominations at prestigious conferences.
\end{IEEEbiography}

\end{document}